\documentclass[sigconf]{acmart}
\usepackage{tabularx}
\usepackage[table]{xcolor}
\AtBeginDocument{%
  }

\copyrightyear{2026}
\acmYear{2026}
\setcopyright{cc}
\setcctype{by}
\acmConference[Koli Calling '26]{26th Koli Calling International Conference on Computing Education Research}{November 5--8, 2026}{Koli, Finland}

\acmISBN{978-1-4503-XXXX-X/2018/06}

\begin{document}

\title[Flawed but Memorable]{Flawed but Memorable: Student Critical Reception of Interest-Personalized GenAI Analogies in Computing Education}


\author{Seth Bernstein}
\affiliation{%
  \institution{University of Michigan}
  \city{Ann Arbor}
  \country{United States}}
  \orcid{0000-0002-7552-5448}
\email{sethbern@umich.edu}

\author{Naaz Sibia}
\affiliation{%
  \institution{University of Toronto}
  \city{Toronto}
  \country{Canada}}
  \orcid{0000-0001-7628-7077}
\email{naaz.sibia@utoronto.ca}

\renewcommand{\shortauthors}{Bernstein and Sibia}

\begin{abstract}
\textbf{Motivation:} Undergraduate computing students increasingly turn to generative AI (GenAI) tools to understand abstract concepts, sometimes receiving explanations built around analogies. Analogies work by comparing an unfamiliar concept to something familiar, but this creates a trap. To tell whether the comparison actually holds, a student needs to know both the concept and the everyday thing it is compared to, and a learner may be missing either one. GenAI may also embed assumptions about who the learner is. Research on GenAI in computing education centers on output correctness at the code and explanation level, leaving students' critical reception of analogies largely unexamined.\\
\textbf{Method:} We investigate how undergraduate students evaluate the accuracy, appropriateness, and embedded assumptions of GenAI-generated analogies, and how they perceive interest-personalized analogies relative to generic technical explanations. Ten students who had completed CS2 participated in this study, with each finishing a pre-survey, a think-aloud annotation task with personalized analogies and generic explanations for linked lists and recursion, and a semi-structured interview grounded in the Paul-Elder critical thinking framework. Students judged analogies along separable dimensions, treating accuracy, clarity, engagement, and trust as distinct standards.\\
\textbf{Results:} Most participants described the interest-personalized analogies as more engaging or memorable than the generic technical explanations, while their accounts of trust were mixed. Some reported trusting the tailored analogies more; others scrutinized them more closely or distrusted the tailoring itself. Participants who demonstrated deep knowledge of an analogy's source domain identified structural flaws that required source-domain knowledge to recognize. Because personalization and explanation format differed together, these findings do not isolate an effect of personalization alone.\\
\textbf{Implications:} A familiar source flips the student's role. On the concept they are still \textit{learners}, but on the familiar source it is compared to they are the \textit{expert}, and that is the position from which an analogy can be judged. We call this two-sided analogy auditing. GenAI systems should therefore ask what students know, not just what interests them, and treat a flawed analogy as something to inspect and fix rather than accept.
\end{abstract}




\maketitle
\section{Introduction}
Generative AI (GenAI) tools have seen broad adoption in computing education, with students using them for tasks ranging from code generation to concept explanation~\cite{prather_beyond_2025, prather_widening_2024}. One documented use is requesting explanations of abstract programming concepts~\cite{macneil_generating_2022}, which LLMs readily deliver as analogies: models produce diverse analogical explanations that students find engaging~\cite{macneil_generating_2022, macneil_experiences_2023}, and both students and instructors now use them to generate personally relevant analogies on demand~\cite{bernstein_analyzing_2024, bernstein_like_2024}. When a student uses GenAI to explain a linked list, the response may arrive as a comparison to everyday objects or experiences. These comparisons are pedagogically appealing because they map unfamiliar structures onto familiar experience, reducing the cognitive load of initial concept formation~\cite{gentner_structure-mapping_1983}. As analogies become a routine mode of AI explanation, understanding of how students receive and evaluate them has not kept pace.

Analogies carry two risks specific to computing education. Because analogies work through structural correspondence between a \emph{source domain} (e.g., Sudoku or video editing) and a \emph{target domain} (e.g., linked lists or recursion)~\cite{gentner_structure-mapping_1983}, a student encountering a concept for the first time may lack the grounding to detect where the mapping breaks down. A linked list analogy that works for traversal may unintentionally misrepresent concepts such as insertion or deletion. Analogies may also encode assumptions about whose experiences count as familiar. For students whose identities or backgrounds do not match the implied learner, those assumptions can signal that they do not belong in the field~\cite{lewis2017fitting, master2016computing}.

Research on GenAI in computing education often concentrates on code generation and technical explanation accuracy~\cite{bernstein_beyond_2025}, leaving students' critical reception of analogies largely unexamined. Work on analogy in computing education addresses structural properties and common misconceptions~\cite{halasz1982analogy,clancy2005misconceptions} but does not ask how students evaluate the analogies they receive, what criteria they apply, or whether they notice embedded assumptions. Interest-based personalization offers one approach to making analogies more personally meaningful~\cite{bernacki_role_2018}. In this paper, \textit{interest-personalized} means that a participant's self-reported interest was supplied to the model as the requested source domain for an analogy; it does not refer to adaptation based on a broader learner model. Whether students perceive or evaluate these interest-personalized analogies differently from generic technical explanations remains unknown. To investigate this question, we conducted a think-aloud annotation task and semi-structured interview with undergraduate students who had completed CS2, grounding the interview protocol in the Paul-Elder critical thinking framework~\cite{paul2006critical}, a widely used model in critical thinking education that evaluates reasoning against a set of intellectual standards rather than treating an explanation as simply correct or incorrect. The framework provided prompts around standards such as accuracy, clarity, assumptions, and point of view, letting us examine not only whether students accepted an analogy but how they justified those judgments. 

Prior work suggests that GenAI tools tend toward surface-level comparisons rather than deep structural mappings~\cite{logacheva_evaluating_2024}, making analogy evaluation a productive site for studying critical thinking because the structural gaps are often invisible to novice learners. Throughout, we distinguish three things a student may do with an analogy: judge its \textit{accuracy} (whether the mapping is structurally correct), judge its \textit{appropriateness} (whether it suits them as a learner, in tone, assumed knowledge, and cultural reference), and decide how much to \textit{trust} it (how far to rely on it without further checking). We use \textit{scrutiny} for the effort a student puts into these judgments and \textit{reception} for their overall response, including engagement and preference. 

Our findings position analogy evaluation as a two-sided knowledge problem: students drew on knowledge of the computing concept to identify which relations an analogy must preserve, and on knowledge of the source domain to check whether those relations held there. Our findings suggest that selecting a source domain the student knows well may position that student to audit the source side of the mapping. This paper makes two contributions. Empirically, it characterizes the criteria students used to evaluate interest-personalized GenAI analogies and generic technical explanations, including structural accuracy, clarity, trust, and assumptions about the learner. Conceptually, it introduces two-sided analogy auditing to explain how target-domain and source-domain knowledge jointly support evaluation of an analogical mapping. We address three research questions:

\begin{itemize}
    \item[\textbf{RQ1}] What criteria do undergraduate computing students apply when evaluating the accuracy and appropriateness of GenAI-generated explanations?
    \item[\textbf{RQ2}] What affective and social dimensions do students perceive in GenAI-generated explanations, including assumptions about learner identity and tone?
    \item[\textbf{RQ3}] What differences do students perceive between interest-personalized analogies and generic technical explanations in engagement, trust, and preference?
\end{itemize}

\begin{figure}[t]
    \centering
    \includegraphics[width=\columnwidth]{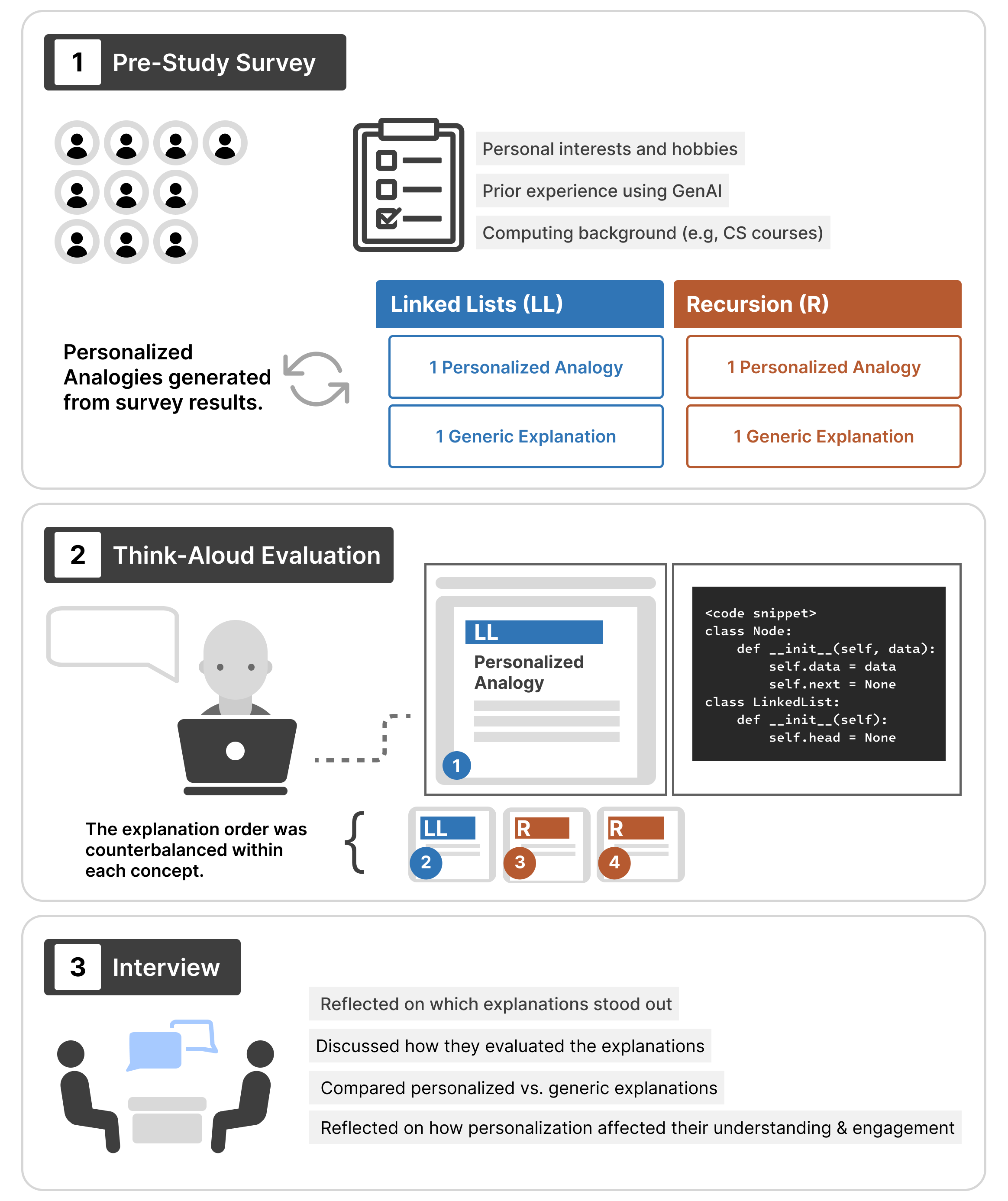}
    \caption{Study procedure. A pre-study survey informed the personalized analogies. Participants then evaluated four explanations (personalized and generic for linked lists and recursion), with explanation order counterbalanced within each concept, in a think-aloud task, followed by a semi-structured interview that opened with general background before turning to the study materials.}
    \label{fig:study-flow}
\end{figure}

\section{Related Work}
\subsection{Analogies in Computing Education}
Analogies do more than simplify abstract concepts. They can function as \textit{constraining representations} that shape how learners interpret unfamiliar material by anchoring it to prior knowledge~\cite{ainsworth_functions_1999}. In computing education specifically, this constraining function is double-edged. Halasz and Moran~\cite{halasz1982analogy} argued that analogical models are structurally partial, requiring learners to distinguish relevant from irrelevant inferences without yet possessing the domain knowledge to do so reliably. Clancy's survey of programming misconceptions~\cite{clancy2005misconceptions} documents how analogies can actively mislead. The common ``variable as box'' metaphor, for instance, implies that a variable might hold more than one value simultaneously. These risks are amplified at the CS2 level, where concepts such as linked lists and binary search trees involve pointer semantics and recursive structure that few everyday analogies capture faithfully. A well-designed analogy bridges the gap between a learner's \textit{concept model} and their \textit{program model}~\cite{heinonen2023synthesizing, sorva2013review}, while a poorly designed one widens that gap. Bettin and Ott~\cite{bettin_pedagogical_2023} formalize this requirement as domain isomorphism, arguing that analogies for computing concepts should preserve the structural relations of the target domain, not only its surface features.  

Even well-designed analogies are not received uniformly. Preference for a given analogy varies across students, and students do not always apply the same analogy the same way~\cite{bettin2022semaphore, bettin2023more}. When students construct their own analogies rather than evaluating provided ones, they face a related difficulty in achieving complete structural alignment while maintaining an appropriate level of abstraction~\cite{harper2023investigating}. Even when an instructor supplies the analogy directly, its measurable benefit is uneven. In a controlled study alternating analogy and no-analogy instruction across CS1.5 topics, Cao et al.~\cite{cao2016examining} found that a brief instructor-presented analogy yielded a short-term performance gain that was statistically significant only for recursion, did not persist into later exams, and did not reliably support transfer to non-technical scenarios. The effect also varied by population: non-majors, first-generation, and international students benefited more from the recursion analogy, whereas CS majors sometimes performed worse, particularly when asked to map a non-technical analogy back onto a technical task. In algorithms courses, students who draw structural comparisons to previously seen problems, rather than superficial ones, tend to score higher on related exam questions, suggesting analogical reasoning is closely tied to problem-solving success even when left implicit in the curriculum~\cite{liu_analogical_2026}.

\subsection{GenAI-Generated Explanations}
GenAI tools can generate code explanations that students rate as more accurate and easier to understand than explanations written by their peers~\cite{sarsa_automatic_2022, leinonen_comparing_2023}, and evaluations of GenAI-generated explanations for logical errors show usable but uneven quality~\cite{balse_evaluating_2023}. Recent empirical work examines how students perceive different configurations of GenAI-generated explanations~\cite{macneil_experiences_2023} and how students and teachers compare AI-generated concept explanations against those produced by human instructors. Lee and Song~\cite{lee_teachers_2024} found that elementary school teachers and sixth-grade students applied similar evaluative criteria of clarity, relatability, and pedagogical effectiveness but diverged in preference. Teachers favored AI-generated explanations for simpler sequential and conditional concepts and human-created explanations for structurally complex ones such as iteration. Both groups struggled to identify whether an explanation was AI- or human-generated, and students struggled more than teachers.

Liu and M'Hiri~\cite{liu_beyond_2024} developed a GPT-3.5-powered virtual teaching assistant for an introductory Python course. The virtual TA matched human TAs on response accuracy for general conceptual questions and outperformed them on clarity and engagement ratings, yet proved less reliable for assignment-specific questions, where human TAs drew on contextual knowledge of course expectations.

Explanation format shapes engagement as well. Hallal et al.~\cite{hallal2026beyond} found that undergraduate chemistry students rated AI-generated stories and analogies as more motivating, enjoyable, and memorable than expository text, with over 80\% reporting that AI-generated narratives improved conceptual clarity. Students preferred all formats combined over stories or analogies alone, positioning narrative AI content as a complement to conventional instruction.

\subsection{Critical Evaluation and Personalization}
GenAI tools acting as on-demand explanation engines raise questions about who controls analogy selection and what assumptions the generated analogies embed. Prior work on student reception of AI-generated content has largely focused on factual accuracy~\cite{balse_evaluating_2023, smith_evaluating_2024}, leaving the social and pedagogical dimensions of AI explanations underexamined. We draw on the Paul-Elder critical thinking framework~\cite{paul2006critical}, which foregrounds the analysis of assumptions, point of view, and fairness alongside accuracy and clarity. The framework lets us ask whether a GenAI analogy is appropriate as well as correct: whose world it assumes, who it positions as a competent learner, and whether its tone and framing could support or undermine a learner's sense of inclusion~\cite{lewis2017fitting, master2016computing}. Empirical work using this same framework bears out the concern that GenAI output may not be received critically. Cardell-Oliver et al.~\cite{cardell2026llm}, analyzing LLM-led Socratic dialogues on computing ethics with the Paul-Elder question taxonomy, found that models scaffolded a broad range of critical-thinking questions but also praised underdeveloped or incorrect responses without due cause, reformulated students' answers rather than prompting students to do so, and rarely challenged shallow or off-topic contributions. Read through Krakauer's~\cite{krakauer2016will} distinction between competitive and complementary technology, such behavior risks performing a learner's cognitive work rather than eliciting it; some students in their study did not engage critically with the tool at all. Whether a GenAI explanation invites scrutiny thus depends not only on the learner but on interactional tendencies of the model itself, such as unearned affirmation.

Interest-based personalization adds a further dimension. Contextualizing computing instruction around student interests has a long lineage, from media computation curricula~\cite{guzdial_exploring_2013} to courses built around creative expression~\cite{guzdial_designing_2025}. Personalizing instruction to individual interests can increase engagement and performance, with effects moderated by how deeply students engage with their interest areas~\cite{walkington_personalizing_2019, bernacki_role_2018}, and interest development theory ties these effects to the transition from situational to individual interest~\cite{hidi_four-phase_2006}. In computing education, personalization and adaptation efforts have so far targeted practice materials such as Parsons problems, worked examples, and contextualized exercises~\cite{ericson_investigating_2019, hou_personalized_2025, pitts_personalized_2026, del_carpio_gutierrez_evaluating_2024} rather than conceptual analogies. Work on motivated reasoning~\cite{kunda1990case} suggests personally relevant material may reduce critical scrutiny, raising the question of whether personalization enhances reception or bypasses evaluation. Studies of student trust in GenAI find that trust varies with task, prior experience, and perceived competence rather than holding steady across contexts~\cite{amoozadeh_trust_2024}. Whether students evaluate personalized analogies differently from generic explanations remains open. More specifically, prior work does not explain how personalization changes the knowledge learners can bring to evaluating an analogy's source domain and its correspondence with the target concept. This study addresses that gap.

\section{Method}
\label{sec:method}
\subsection{Study Design}
Each participant evaluated personalized analogies and generic explanations for linked lists and recursion. This supported within-participant comparison, while reflexive thematic analysis examined patterns across the dataset. We recruited students who had completed CS2 so that they could evaluate analogies against concepts they understood, isolating their critical judgment of the analogy from the separate difficulty of learning the concept itself. We selected these concepts because they are foundational CS2 topics that students commonly find challenging; prior work identified recursion as a threshold concept in computing education~\cite{sanders2016threshold,boustedt2007threshold}, and students in Layman et al.'s CS2 study ranked linked lists among the most challenging topics~\cite{layman2020toward}. For each concept, participants saw one personalized analogy drawn from their stated interests and one generic explanation generated with no theme specified. Every participant completed the linked list concept first, followed by recursion; this concept order was fixed across all participants. Within each concept, the order of the personalized and generic explanations was counterbalanced (Section~\ref{sec:thinkaloud}). The two generic explanations were identical for every participant, while each participant's personalized analogies were generated from their own interests. The study combined a think-aloud annotation task with a semi-structured interview. 

This design captured behavioral evidence of evaluation criteria through annotation artifacts and reflective verbal data through the interview, holding the concepts, learning objectives, and generic explanations constant across participants while each participant's personalized analogies were drawn from their own interests. The personalized condition was always an analogy and the generic condition a technical explanation, so the two conditions differ in format as well as personalization; we return to what this permits and precludes in Section~\ref{sec:limitations}. Figure~\ref{fig:study-flow} summarizes the full procedure.

\subsection{Participants}
We emailed 667 students who had completed CS2 at a large public research university in North America (University of Toronto), and from this pool recruited ten participants. Participation was voluntary, and each session was compensated with a \$30 gift card. Sessions ran individually via Zoom conducted by one of two interviewers, lasted approximately 60 to 90 minutes, and were audio and video recorded with participant consent. Participants were assigned to the two counterbalanced presentation orders by participant number, yielding equal group sizes. This study was approved by the Institutional Review Board (Protocol \#47002). Table~\ref{tab:participants} lists each participant's intended major, year in the program, and the interest domains used to generate their personalized analogies.

\begin{table}[t]
\centering
\small
\begin{tabularx}{\columnwidth}{@{}cc>{\raggedright\arraybackslash}X>{\raggedright\arraybackslash}X@{}}
\toprule
\textbf{ID} & \textbf{Yr} & \textbf{Intended Major} & \textbf{Interests} \\
\midrule
\rowcolors{2}{white}{gray!12}
P1  & 3 & Computer Science and Statistics & Soccer; Avengers (film) \\
P2  & 2 & Information Security and Computer Science & DaVinci Resolve (video editing); Valorant (game) \\
P3  & 2 & Computer Science & Piano; Valorant (game) \\
P4  & 3 & Computer Science and Statistics & Karate; Piano \\
P5  & 1 & Computer Science & The Last of Us (game); Chess \\
P6  & 1 & Computer Science and Applied Statistics & One Piece (manga/anime); Badminton \\
P7  & 1 & Applied Statistics & Minecraft (game); Gym \\
P8  & 3 & Statistics & LEGO; Sudoku \\
P9  & 2 & Computer Science and Applied Mathematics & Painting; Colour mixing \\
P10 & 2 & Computer Science & Baking; Painting \\
\bottomrule
\end{tabularx}
\caption{Participant year, major, and interest domains used for personalized analogies. Odd-numbered participants were assigned to Condition A, even-numbered to Condition B.}
\label{tab:participants}
\end{table}

\subsection{Pre-Survey}
Participants completed a pre-survey before their session. The survey collected computing coursework history, personal interests, prior experiences with analogies and GenAI tools, and sense of belonging in computing, measured with three items adapted from established belonging scales~\cite{good2012belonging}. The interests section prompted participants to be specific, asking for hobbies, media, or domains they engaged with regularly rather than broad categories. Researchers used these responses directly for analogy generation.

\subsection{Analogy Generation}
Researchers generated one interest-personalized analogy per concept by supplying a participant's self-reported interest as the requested source domain. When participants listed multiple interests, one researcher selected pairings based on structural fit. Imperfect mappings remained suitable because participants were asked to critique and repair them. A consistent prompt structure specified the learning goals for each concept and incorporated the student's stated interest as the analogical source domain. For linked lists, the goals were element storage, node references, and traversal. For recursion, they were self-reference, problem reduction, and the base case. The recursion prompt read: \textit{``You are explaining recursion, specifically the factorial function, to a CS2 student working in Python 3.11 who is interested in [interest]. Use an analogy from their interest area\ldots{} Cover: how a function calls itself, how each call reduces the problem, and how the base case stops the process. Write in plain paragraphs only\ldots{} under 300 words.''} The linked-list prompt followed the same structure, substituting element storage, node references, and traversal as the goals. Only the bracketed interest and target concept varied across participants; the tone instruction, length limit, and format constraints were identical. Two generic explanations, one per concept, were generated once and held constant across all participants, produced with the same prompt structure but with no theme specified. Personalized and generic explanations shared an identical prompt structure and learning objectives; the personalized prompt additionally specified the participant's stated interest as the source domain for an analogy, while the generic prompt requested a technical explanation with no theme. 

Before data collection, researchers generated candidate explanations using OpenAI's GPT-5.5 (Thinking: High) and Anthropic's Claude Opus 4.8 (Thinking: High), and for each explanation one researcher selected the output that best covered the stated learning goals and read clearly. We treated the two models as interchangeable sources of candidate explanations rather than as objects of comparison, and did not track which model produced each selected output.

Researchers did not modify outputs or deliberately introduce structural errors, and structural fidelity was not a selection criterion, so any structural imperfections in the personalized analogies arose from the generation process rather than from researcher edits. Researchers checked that each explanation covered the intended learning goals and read clearly, but did not require the analogies to be structurally perfect, since identifying and repairing imperfect mappings was itself part of the task participants were asked to perform. 

\begin{figure*}[t]
    \centering
    \includegraphics[width=\textwidth]{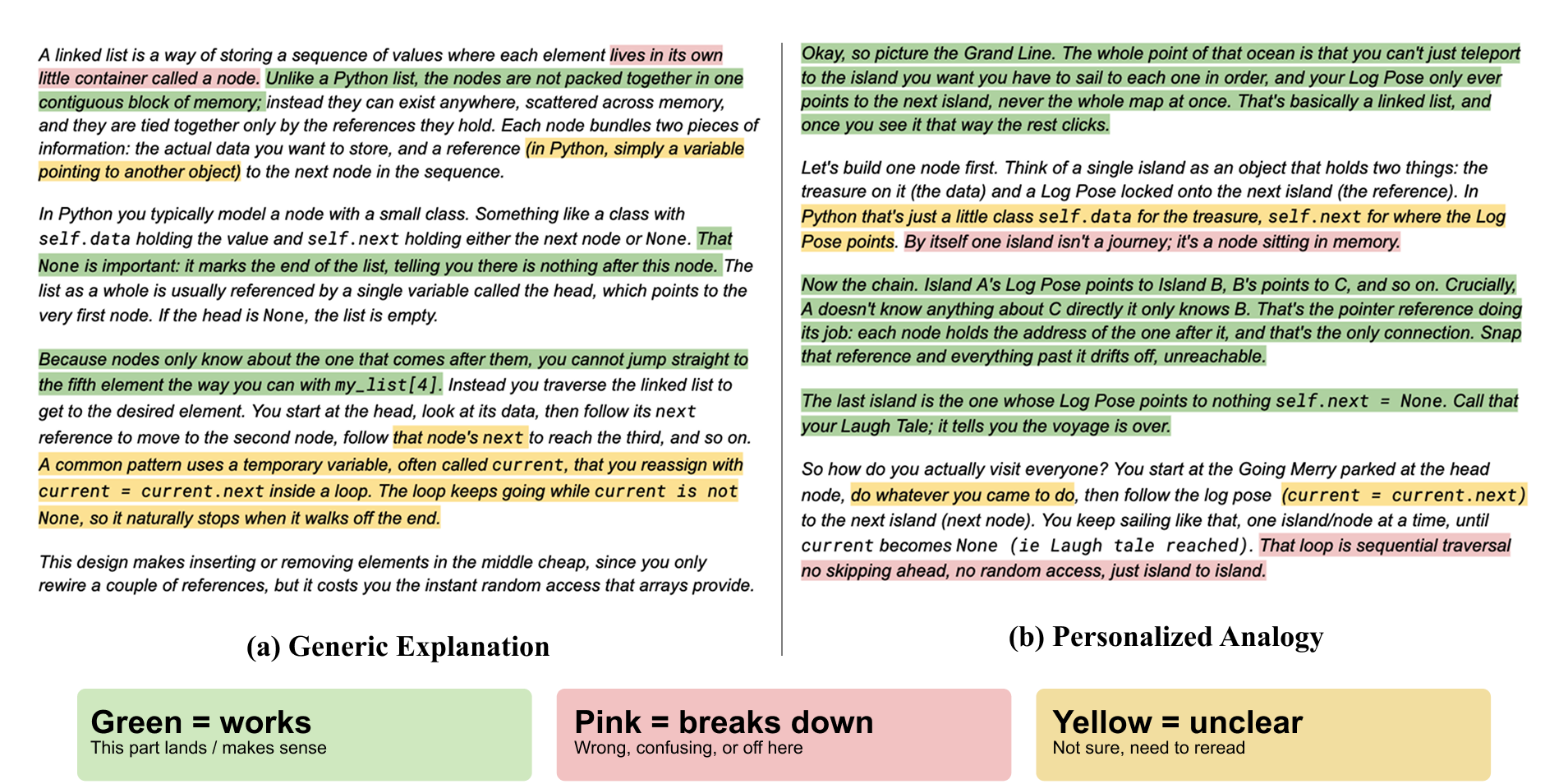}
    \caption{Generic (left) and personalized (right) linked list explanations from the study. The personalized version incorporates a participant's stated interest in \textit{One Piece}, a long-running manga and anime series. P6 later identified a structural mismatch in this mapping: the island route loops back to its start, making the source domain closer to a circular linked list than the singly linked list under study.}
    \label{fig:linked-list-explanations}
\end{figure*}

The resulting stimuli varied in how faithfully the source domain preserved the concept's structure. Several recursion analogies mapped cleanly, since the source domain guaranteed a shrinking input toward a base case, as in an Avengers mission that hands off a smaller task at each step or a Sudoku puzzle reduced one cell at a time. Others fit loosely, since the source carried no guaranteed reduction: a badminton rally has no fixed length, and participants flagged exactly these looser mappings. One linked-list analogy built on a circular route, closer to a circular linked list than the singly linked list under study. This procedure yielded stimuli with a range of structural fidelity, giving participants both stronger and weaker mappings to evaluate. Each concept was paired with a fixed code snippet held constant across participants. Figure~\ref{fig:linked-list-explanations} shows a generic and a personalized linked list explanation side by side.

\subsection{Think-Aloud Task}
\label{sec:thinkaloud}
Each participant evaluated all four explanations one at a time in a Google doc created for them, shared with the interviewer, chosen so students could annotate and edit in an environment they already knew, with each explanation presented alongside the relevant code snippet. Participants used the document as a live workspace, annotating and editing directly as they thought aloud. Presentation order was counterbalanced across two conditions to control for ordering effects within each concept. In Condition A, participants saw the generic explanation before the personalized analogy for each concept; in Condition B, this order was reversed. Odd-numbered participants were assigned to Condition A and even-numbered participants to Condition B. Before the task began, participants completed a brief think-aloud warm-up to familiarize themselves with verbalizing their reasoning.

For each explanation, participants received the following prompt: \textit{``React to this analogy/explanation, evaluate it, and suggest any fixes or changes you would make. Verbalize everything as you go.''} The interviewer stayed silent except for neutral prompts such as ``keep talking'' or ``tell me more''. Within the Google Doc, participants highlighted and labeled text, edited it directly, and added or deleted content. Annotation artifacts were preserved for analysis.

\subsection{Interview}
Following the think-aloud task, participants completed a semi-structured interview organized into three parts. Part one asked about computing background and prior experiences with explanations in general, independent of the study materials: what kinds of explanations work best for them, whether analogies help more than technical definitions, and whether an explanation had ever felt confusing or like it was designed for someone else. Part two turned to retrospective reflection on the think-aloud, opening with a bridging question about which explanation stuck with them most, then moving to probes derived from the Paul-Elder framework~\cite{paul2006critical}: whether participants doubted the accuracy of any explanation and how they would verify it (accuracy), which explanations were hardest to follow and why (clarity), what each explanation assumed about their prior knowledge (assumptions), and whose perspective each explanation seemed written from (point of view). Part three asked participants to compare the two explanation types, reflect on whether personalization changed how critically they read, and offer recommendations for instructors and tool designers. 

\subsection{Analysis}
Our analysis operates within an interpretivist epistemology, following the reflexive (Big Q) orientation to thematic analysis rather than a coding-reliability or codebook orientation \cite{braun2023doing}, consistent with recent guidance for thematic analysis in computing education research that cautions against importing positivist criteria such as intercoder reliability or saturation into interpretivist work~\cite{padiyath2026reflecting}. We treat participant accounts as situated interpretations shaped by the interview context, prior coursework, and individual disposition toward GenAI tools rather than as neutral reports of an underlying evaluation process. Reflexive thematic analysis fits this stance, since it treats the researcher's own positionality as part of the analytic process~\cite{braun_using_2006}. Both analysts are computing education researchers who study GenAI-based learning tools and approached the data with an existing interest in personalization. We treated this orientation as a lens to interrogate during analysis rather than a bias to eliminate. 

We analyzed interview transcripts and think-aloud recordings using this approach. Following familiarization with the recordings and transcripts, both researchers coded the full dataset independently to develop distinct interpretive readings rather than to assess reliability. They then met repeatedly to compare interpretations, interrogate assumptions, and construct candidate themes, refining those themes against the transcripts and annotation artifacts. We did not calculate intercoder agreement or reconcile all codes into a consensus codebook. Differences between readings served as analytic resources in the collaborative construction of themes. Annotation artifacts served as a secondary data source and were examined for patterns in what participants flagged, edited, or deleted across explanation types.

Participant quotes reported below come from session transcripts, lightly edited for readability, with omissions marked by ellipses and clarifications added in brackets. The analysis attended to differences between personalized and generic conditions and to variation across participants. The Paul-Elder framework structured the interview probes but did not serve as a coding frame; themes were constructed inductively from the data, and we read the framework's standards against those themes during interpretation rather than mapping data onto them. Where we report that a pattern held for ``most'' or ``nearly all'' participants, these are descriptions of its spread within this sample of ten rather than claims about prevalence in a wider population. Table~\ref{tab:codebook} summarizes the themes we constructed, provided to orient the reader rather than as a coding instrument applied for reliability. Because we organize the findings by research question rather than by theme, these themes surface across sections: structural correspondence and precision of language under RQ1, assumed background and audience under RQ2, and trust, engagement, and control under RQ3.

\begin{table}[t]
\centering
\small
\renewcommand{\arraystretch}{1.25}
\begin{tabularx}{\columnwidth}{@{}>{\bfseries\raggedright\arraybackslash}p{0.33\columnwidth}X@{}}
\toprule
Theme & What it captures \\
\midrule
Learning preferences & Students' initial approaches to understanding unfamiliar CS concepts before using AI. \\
Personalized analogies & Reactions to explanations tailored to participants' stated interests. \\
Fairness \& assumptions & Perceived assumptions about learners' backgrounds, prior knowledge, and tone. \\
Trust \& verification & How students judged the credibility of explanations and when they chose to verify them. \\
Evaluation criteria & The qualities participants used to assess explanation quality (e.g., clarity, accuracy, usefulness). \\
Design recommendations & Participants' suggestions for improving AI-generated explanations and analogies. \\
\bottomrule
\end{tabularx}
\caption{Top-level themes constructed through reflexive thematic analysis. Each theme encompasses multiple lower-level codes.}
\label{tab:codebook}
\end{table}

\section{Results}
We present findings organized by research question, drawing on the themes generated through the reflexive thematic analysis described in Section~\ref{sec:method}. The pre-survey served to generate each participant's personalized analogies and to characterize the sample rather than as a source of analyzed outcomes, so the findings below rest on the think-aloud and interview data. Asked about explanations in general, participants described when analogies help them. Nearly all participants saw analogies as most useful at the start of learning a concept, for an initial big-picture understanding, with technical detail more appropriate once that understanding had formed. As P9 put it, ``when I'm just introduced to one, then analogies help\ldots{} and then once I get [deeper]'' the technical explanation takes over. Across all three questions, participants treated the analogies as arguments with a point of view rather than as true-or-false statements. They checked their structural accuracy, examined whose experience they assumed, and weighed the relationship personalization presumed with the learner. Across these judgments, participants drew on knowledge of both the computing concept and the analogy's source domain, a pattern we develop in the Discussion as two-sided analogy auditing.

\subsection{Evaluation Criteria (RQ1)}
Participants did not evaluate analogies holistically. Across the open-ended think-aloud task and the interview probes, they flagged specific, low-level breakdowns in the mapping between source and target domain. A recurring failure participants identified was a structural mismatch between the analogy and the real-world domain it borrowed from, echoing Halasz and Moran's~\cite{halasz1982analogy} concern that analogical models are only ever partially structure-preserving. 

P6, evaluating a linked-list analogy built around \textit{One Piece} (a manga series in which characters sail between islands in a fixed sequence), corrected the analogy's structure rather than the computing concept itself: ``if you really consider [it] to be a link[ed] list, it would be a circular link[ed] list, because once you pass the last island, you go back to the first island.'' The analogy implied a loop back to the start, which describes a circular linked list rather than the singly linked list under study. P8 raised a similar objection to a Sudoku-based recursion analogy, noting that ``having a base case of 1 in this analogy is a slight disconnect between Sudoku and recursion\ldots{} you can't always just write one for that one remaining cell.'' P2 identified a subtler agency error in a video-editing analogy for linked lists, observing that the analogy misattributed the traversal logic: ``it's more of like a case where the playhead knows what's next, rather than, like, the clip knows what's next.'' P6 rejected a badminton-based recursion analogy on the same grounds, since a rally carries no guarantee of a shrinking input, ``if I had to compare a badminton rally to anything, it would be a loop, not a recursion,'' and proposed golf, where each successive stroke covers less distance, as a structurally sound replacement. P3 rejected a teammate-handoff analogy for recursion because the tasks it described lack the dependence recursive calls require: ``they're really just, like, independent when we sync up tasks, so I would say, like, this is not that good of an example.'' When students had enough domain expertise to check an analogy's mechanics, they held it to whether its structure matched the concept, not only whether it captured the general idea.

A second recurring criterion was precision of language. Participants flagged specific phrases as ``open to interpretation'' rather than critiquing tone or content broadly. P8 pointed to hedging language directly: ``using slightly simpler answer isn't really telling us much\ldots{} that is a little bit open to interpretation, which you don't want, obviously.'' P1 made a concrete substitution proposal rather than a vague complaint, suggesting an explanation ``could have said that it points to the memory address of the next node'' instead of the vaguer ``remembering.'' Related to this, several participants flagged \textit{missing} elaboration as a distinct failure from vagueness: P4 wanted ``an additional thing on why it's necessary to have a temporary variable instead of using, like, directly the one,'' and P7 argued that fairness required the AI to justify claims rather than assert them: ``it should be more specific. This is here for this reason.'' Clarity judgments also bled into accuracy judgments. P5, having just described an explanation as accurate, doubted that accuracy on the grounds that the personalized text was ``a lot more bloated'' than the alternative.

Participants also described concrete verification behaviors rather than passive acceptance, which speaks to how these criteria were applied in practice rather than only in the abstract. P4 described cross-referencing AI output against lecture material as a default habit: ``I would always either fact check it, like, based on the course notes\ldots{} or ask it questions so it kind of sits right, like, based on what I've learned in lecture.'' P5 made trust conditional on provenance: ``I usually try my best to give it a source of information\ldots{} I would be a lot more doubtful and untrustworthy [without one].'' Rather than a single accept-or-reject judgment, participants' evaluation worked as an iterative filter, with follow-up questions used to probe for inconsistency. P6 described testing analogies against worked examples ``so that I can put them in and see what the expected result is.''

\subsection{Affective and Social Dimensions (RQ2)}
\label{sec:rq2}
Participants also evaluated explanations for whose background they assumed, echoing the fairness and point-of-view standards of the Paul-Elder framework~\cite{paul2006critical}. The most direct failure mode was assuming prior knowledge the student did not have yet. P1 described being confused by an explanation that reached beyond the course syllabus: ``it gave a lot of theorems and formulas, which I wasn't\ldots{} I haven't learned before.'' P8 reported a similar mismatch when practice questions assumed upper-year coursework. P7 noted that this gap was structural to how the AI scaffolds new material, jumping from foundational data structures to more advanced ones ``with no background whatsoever.''

A related failure was assumed domain familiarity, particularly around sports and gaming references. Asked in the interview whether an explanation had ever felt like it was not meant for them, P9 recalled past experiences with analogies drawn from ``the more male-dominated\ldots{} [side] of computing\ldots{} I personally don't know much about sports, so using those analogies, I'm not sure what to do with that.'' The concern was about the kinds of references AI explanations reach for by default rather than about a specific analogy in the study. P2, an avid gamer evaluating a recursion analogy built around \textit{Valorant} (a team-based competitive video game), flagged the same failure mode even for a game they knew well: ``it does assume that you have the terminology you need\ldots{} if you had somebody who doesn't play Valorant\ldots{} they might be thinking, like, what's an entry fragger [a specialized in-game role]?'' P1 raised this as an explicit design recommendation (Section~\ref{sec:rq3}), warning designers against analogies drawn from ``movies which they haven't watched\ldots{} or sports they've never played before.''

Participants also read excessive explanation as a social signal. When an explanation over-explained familiar material, several took it as a sign the tool assumed they knew less than they did. Participants described excessive explanation as condescending rather than inefficient. P9 was the most explicit: ``it felt like it was coming from a point of view where it was trying to baby me, in a sense\ldots{} it assumed I knew less than I actually do, and then tried to make up for that.'' P6 described the same phenomenon in terms of word economy: ``I feel like they've picked, like, a really complicated way of explaining this\ldots{} at least 5 too many words than there should have been.'' This suggests students read \textit{length} and \textit{repetition} as social signals of assumed competence, not just as stylistic choices, complicating a simple ``more scaffolding is safer'' design heuristic.

Not all participants attributed confusion to the explanation itself. P10 explicitly declined to locate any fairness failure in the AI's assumptions, instead self-attributing confusion to their own effort: ``maybe I'm not understanding this right now, but I need to put in more effort.'' We report this as a distinct pattern rather than an absence of finding, since it indicates that not all students apply the same fairness lens by default, and some may locate difficulty in their own effort rather than interrogating the explanation's assumptions.

\subsection{Personalization and Reception (RQ3)}
\label{sec:rq3}
Participants reported sharply divided reactions to the interest-personalized analogies, and those reactions were often associated with whether they knew the analogy's source domain well. When personalization ``landed,'' participants described a qualitative shift in how memorable or intuitive a concept became, consistent with Hallal et al.'s~\cite{hallal2026beyond} finding that AI-generated analogies increase perceived engagement. P6 described a durable shift in how they now think about linked lists: ``I'm not gonna lie, this might have even helped me to understand linked lists\ldots{} I might in future think about it as a Log Pose [the series' island-to-island compass].'' P8 described personalization as changing their overall stance toward analogies as a category: ``you really changed my mind about analogies\ldots{} definitely the custom, personalized ones were very, very awesome.'' P9 attributed the analogy's usefulness specifically to lived experience with the source domain, a painting analogy that worked ``because it connected to something I am quite familiar with\ldots{} since I've gone through those motions myself.''

When the fit was poor, participants went further, calling the personalization artificial. P10, whose stated interests included baking, still rejected the baking-based analogy because it would not generalize: ``I didn't really like this analogy. I feel like not everyone bakes.'' The objection concerned generalizability rather than personal fit; an interest-based analogy risks excluding students who do not share the interest. P7 described a gym-based recursion analogy as reaching for a connection that was not structurally there: ``it's digging too much into their interests, it's trying to find something that isn't there.'' Several participants preferred the plain technical explanation \textit{even when the personalized one matched their own stated interest}. P7, evaluating a \textit{Minecraft}-themed explanation (a sandbox building game), said: ``even though Minecraft was my bias, I actually like this one more [the generic one], because it feels more, like, technical.'' Asked why, P7 explained that the personalized analogy, while relatable, felt incomplete: ``even though it's, like, nice to relate to, I wouldn't just rely on that, I would ask an additional prompt to go more into detail, whereas this one is\ldots{} a good, concise introduction to the concept.'' P7 preferred the technical explanation for its self-sufficiency since it stood on its own, while the analogy would have required follow-up to reach the same depth. Interest matching alone did not guarantee reception when the personalized version sacrificed completeness for relatability.

The clearest evidence bearing on motivated reasoning~\cite{kunda1990case} came from participants' own accounts of whether personalization changed their critical stance. The pattern was not uniform. P1 reported exactly the reduction in scrutiny our framework anticipated, saying that knowing the analogy was tailored to them left them ``less critical about it because I could trust it more.'' But P7 reported the opposite association between personal relevance and trust, treating heavy personalization as a reason for suspicion rather than comfort: ``that actually made me trust them less\ldots{} I tried to let AI know the bare minimum about me.'' What tracked scrutiny consistently, across participants, was domain expertise rather than personal interest per se: P9 noted that their ability to audit an analogy came from prior knowledge of the target concept, not affection for the source domain: ``since I knew a lot about the concepts, I could tell if it was going wrong\ldots{} it didn't increase my trust.'' P6 made the same expertise-contingency explicit from the other direction, describing how their fluency in \textit{One Piece} let them appreciate an analogy's craftsmanship in a way a non-fan could not: ``now that me knowing LinkedList and me knowing OnePiece can appreciate that this is a really good explanation.'' Together, these accounts suggest that the relationship between interest personalization and scrutiny may depend less on whether content is interesting than on whether the student has sufficient target- or source-domain knowledge to evaluate it. We treat this as an emergent qualitative interpretation rather than evidence of a causal effect.

Finally, when asked directly how a GenAI tutor should handle personalization, participants converged on giving the learner control rather than having the system infer or impose interests. P1 proposed a menu-based design: ``it should ask the prompter if they would like a real life analogy\ldots{} provide maybe 2 or 3 examples, and then the prompter could choose from those 3 examples.'' P10 proposed a more conservative default, reserving analogy for cases where a technical explanation alone had already failed: ``focus on making a comparison before it dives deeper into the content\ldots{} if the student is unable to understand, then to bring in analogies from the real world.'' Others treated the two forms as complements rather than alternatives. P2 described the personalized analogy as building a big picture that made the generic explanation easier to digest afterward, while P7 proposed the reverse order, a formal definition first with the analogy as reinforcement. This design preference aligns with the fairness findings in Section~\ref{sec:rq2}. Participants who wanted opt-in personalization frequently cited the same risk of alienating peers who did not share the interest, e.g. P1's warning against analogies drawn from ``movies which they haven't watched\ldots{} or sports they've never played before.''

Pre-survey responses were descriptively consistent with this pattern. Participants reporting the lowest confidence in linked lists and recursion tended to raise clarity and wording concerns rather than the structural audits seen from higher-confidence participants. Sense-of-belonging scores, inspected descriptively rather than analyzed as a variable, clustered high across the sample.

\section{Discussion}
Our three questions yielded distinct findings. For RQ1, students evaluated analogies by structural correspondence, holding an analogy to whether its source domain shared the computing concept's structure, and applied precise, mechanism-level criteria rather than global judgments of accuracy. For RQ2, students read explanations for whose knowledge and background they assumed, treating both unfamiliar cultural references and excessive scaffolding as signals about who the explanation was written for. For RQ3, most participants described the personalized analogies as more engaging, while trust moved in no single direction; scrutiny tracked domain expertise more consistently than personal interest. We interpret these findings below.

\subsection{Two-Sided Analogy Auditing}
\label{subsec:two-sided-auditing}
We call our account of how learners evaluate analogies \textit{two-sided analogy auditing}. Target-concept knowledge identifies the relations an analogy must preserve, such as recursion's shrinking input and dependent calls or a linked list's pointer chain. Source-domain knowledge determines whether those relations hold there; neither side is sufficient alone. Structure-mapping theory establishes that analogical reasoning depends on relations between source and target domains~\cite{gentner_structure-mapping_1983}. We show what this may mean for critically evaluating a mapping and propose that a familiar source can affect which knowledge a learner brings to that evaluation.

This reframes how interest personalization may function. Prior literature primarily treats interest-based personalization as a lever for engagement. In our study, participants who knew an analogy's source domain well could also draw on that knowledge when inspecting the mapping. A familiar source may therefore make an analogy both more engaging and more inspectable: familiarity can help a learner identify where a source domain fails to preserve the target concept's structure. We interpret this as a possible epistemic role for interest personalization in addition to its motivational role. Because explanation format and personalization varied together in our design, the study cannot isolate personalization as the cause of the observed scrutiny.

A second implication follows for how the field measures explanation quality. GenAI computing education research tends to focus on quality as a single correctness judgment. Our participants did not evaluate it that way. They assessed the underlying concept, the source domain, and the correspondence between them as separate layers, and accepted one even when they rejected the other, rating analogies as memorable even when they doubted the accuracy, or found it accurate even when objecting to its tone. Constructivist accounts predict this, since understanding is built by integrating new material with the particular knowledge structures a learner already holds \cite{piaget1970science}, and different learners hold different structures. A single quality score would therefore obscure both where an explanation succeeds for a given student, and the layer, concept, source, or mapping, at which a personalized analogy breaks.

Participants' accounts emphasized structural correspondence alongside, and sometimes over, engagement or personal relevance. Alexander's Model of Domain Learning frames the capacity to make that judgment as a function of expertise \cite{alexander_model_2004, alexander_development_2003}. Learners move from acclimation, where they lack the structured knowledge to evaluate claims in a domain, toward proficiency, where a rich knowledge base supports independent judgment. Participants who demonstrated substantial source-domain knowledge audited the source side of the mapping directly, while target-domain knowledge helped them identify which relationships the analogy needed to preserve. These observations suggest that interest and expertise should be distinguished and that expertise should be identified as source-domain or target-domain expertise rather than treated as a single construct. 

We did not see that personalization influenced trust uniformly. Across the ten participants, some grew more suspicious of a familiar analogy, others were trusting while still being critical. What their coexistence suggests is that personalization changes what a learner brings to an evaluation, the structures and dispositions they test new material against, rather than pushing trust one way.

\subsection{Implications for Design and Instruction}
\label{subsec:implications-for-design-and-instruction}
First, we recommend personalizing through source-domain structure rather than topical relevance alone. By \textit{source-domain structure}, we mean relationships and processes in the familiar domain that must correspond to the target concept, such as successive problem reduction and a stopping condition for recursion or a directed chain of references for a singly linked list. Interest matching alone can leave a generic explanation behind. Depth of domain engagement, not surface reference, drove both engagement and auditing, consistent with evidence that personalization depends on how deeply students engage with their interests~\cite{bernacki_role_2018}. Systems should therefore distinguish interest from expertise, asking learners not only what interests them but which domains they know well enough to reason about critically.

Second, verify structure before delivery. Because structural correspondence separated accepted from rejected analogies regardless of relevance, no amount of personal relevance rescues a broken mapping. A generation pipeline could confirm that a candidate source preserves the concept's essential structure before the student sees the analogy.

Third, give the learners control over how personalization is used in their learning, including the option to view a generic explanation alongside a personalized one. 

The most consequential implication is instructional, and it inverts how these systems are usually framed. Locating where a mapping breaks requires knowing the concept's actual structure, so analogy critique is itself a check on understanding, and a more demanding one than reading a finished explanation. A student who shows that a badminton rally lacks a shrinking input has demonstrated an important component of recursion. This is a constructivist claim: critique asks the learner to test an external structure against the target concept they already hold \cite{piaget1970science}, which is the same integrative work that helps build understanding. It reframes the accuracy risk of GenAI analogies as a resource. Rather than filtering flawed analogies out as a hazard, an instructor can assign them as objects to inspect, question, and repair, turning the tool's least reliable property into the substance of the exercise.

\subsection{Motivation or Knowledge}
Our findings add to an open question from prior work. Motivated-reasoning accounts predict that personally relevant material lowers critical scrutiny \cite{kunda1990case}, while the personalization-as-engagement literature predicts it raises reception \cite{walkington_personalizing_2019, bernacki_role_2018}. Both treat personalization as acting on a learner's motivation. Our results located its effect elsewhere. By selecting a domain the learner already knows, personalization changes the knowledge available for evaluation (as we discussed above). 

This distinction clarifies a tension in prior results. Cao et al.~\cite{cao2016examining} found that CS majors sometimes performed worse with an analogy, particularly when mapping a non-technical comparison back onto a technical task, whereas our domain-expert participants audited analogies effectively. The two outcomes measure different knowledge against different demands. Cao's expertise axis is computing knowledge, the target side, assessed against a performance outcome. Ours is interest-domain knowledge, the source side, assessed against an auditing outcome. Which side a learner is strong on determines what they can do with the analogy, so expertise does not raise or lower reception uniformly. It shifts the kind of engagement an analogy encourages.

This also revises a concern about analogical models in computing. Halasz and Moran \cite{halasz1982analogy} argued that learners cannot reliably separate relevant from irrelevant inferences because they lack the domain knowledge to do so. Two-sided auditing shows that this deficit may be specific to source-domain or target-domain knowledge. Where Bettin and Ott~\cite{bettin_pedagogical_2023} treat domain isomorphism as a property designers must guarantee, our participants enforced it themselves, rejecting analogies whose structure diverged from the concept and locating the point of divergence. 

Finally, our account adds a lever that work on GenAI scrutiny has not considered. Cardell-Oliver et al.~\cite{cardell2026llm} locate the failure of critical engagement in the model's interactional habits, such as praising underdeveloped answers and doing the learner's reasoning for them. That places the burden of eliciting scrutiny on how the model behaves. We add that a well-chosen domain equips a student to challenge the explanation regardless of whether the model invites the challenge. Personalizing through source-domain structure is therefore not only an engagement decision but a critical-thinking intervention, one that operates on what the learner can question rather than on what the system chooses to ask.

\subsection{Future Work}
The instructional use of analogy critique remains untested. A classroom study could assign students GenAI analogies with known structural flaws and test whether locating and repairing them predicts performance on an independent assessment, separating critique from exposure to a correct explanation. The study could also distinguish failures identified in the target concept, source domain, or correspondence, directly testing two-sided analogy auditing.

Because our participants had completed CS2, novice evaluation remains untested. Alexander's Model of Domain Learning predicts that novices rely more on surface features and external validation than on independent structural judgment~\cite{alexander_model_2004}. Recruiting students before they encounter linked lists or recursion could test whether source-side and target-side critique depend differently on prior mastery of the interest domain and computing concept. A larger sample could also test whether the observed trust stances correspond to prior GenAI use, self-reported domain expertise, or disposition toward automated systems.

Implementing the structural correspondence check proposed in Section~\ref{subsec:implications-for-design-and-instruction} and comparing reception of checked and unchecked analogies would test whether it improves the outcomes participants described.

\subsection{Limitations}
\label{sec:limitations}
\textbf{Construct validity.} Personalization may be confounded with explanation format. The personalized explanations used analogies, while the generic ones were mainly plain technical explanations. So the engagement differences participants described could come from the analogy format, the personalization, or both. Prior work finds AI-generated analogies more engaging than plain explanations on their own~\cite{hallal2026beyond}, and telling these apart would need a generic-analogy condition. Researchers also curated the stimuli, choosing which interest to pair with each concept and which output to show, so participants saw a filtered set of analogies rather than the raw output a fully automated system would generate.

\textbf{Internal validity.} Because the generic explanations were the same for everyone while the personalized analogies were generated per participant, we cannot separate variation in the personalized stimuli from participants' individual interests. Everyone did linked lists before recursion, so reactions to the recursion analogies may reflect familiarity gained in the first task; we varied the order of personalized and generic versions within each concept, but not the order of the concepts themselves. We collected pre-survey measures of concept confidence and sense of belonging but did not analyze how they related to evaluation behavior, given only ten participants and little variation in belonging scores; the interview data stayed our main source of evidence.

\textbf{External validity.} Ten participants limit how far the findings transfer, and several patterns rest on just one or two accounts. In reflexive thematic analysis a pattern matters for its interpretive weight rather than how often it appears~\cite{padiyath2026reflecting}, so a single clear account can still be a real finding, but claims like P7's privacy discomfort need more study before they inform design. All participants had finished CS2 and evaluated concepts they already knew, so their criteria may differ from those of true beginners. Finally, participants' interests leaned toward gaming, sports, and popular media, the same reference space P9 found alienating, so our stimuli may under-represent how personalization works for students whose interests fall outside it.
\section{Conclusion}
We examined how ten undergraduate computing students evaluated GenAI-generated analogies for linked lists and recursion, comparing interest-personalized analogies with generic technical explanations. Students audited structural correspondence, flagged imprecise language and missing justification, and read assumptions about knowledge and cultural reference points as signals about the imagined reader. Most found the personalized analogies engaging, trust moved in no single direction, and scrutiny aligned more consistently with source- or target-domain knowledge than with personal interest. These findings support tools that verify structural fit, place personalization under student control, and treat analogy critique as a learning activity. A deeply known source domain may help learners challenge the explanation.

\begin{acks}
We thank Hannah Vy Nguyen for designing the study-procedure figure. We also acknowledge the support of the Natural Sciences and Engineering Research Council of Canada (NSERC) PGS D–600673–2025.
\end{acks}

\bibliographystyle{ACM-Reference-Format}
\bibliography{sample-base,seth-zotero}


\begin{thebibliography}{53}


\ifx \showCODEN    \undefined \def \showCODEN     #1{\unskip}     \fi
\ifx \showISBNx    \undefined \def \showISBNx     #1{\unskip}     \fi
\ifx \showISBNxiii \undefined \def \showISBNxiii  #1{\unskip}     \fi
\ifx \showISSN     \undefined \def \showISSN      #1{\unskip}     \fi
\ifx \showLCCN     \undefined \def \showLCCN      #1{\unskip}     \fi
\ifx \shownote     \undefined \def \shownote      #1{#1}          \fi
\ifx \showarticletitle \undefined \def \showarticletitle #1{#1}   \fi
\ifx \showURL      \undefined \def \showURL       {\relax}        \fi
\providecommand\bibfield[2]{#2}
\providecommand\bibinfo[2]{#2}
\providecommand\natexlab[1]{#1}
\providecommand\showeprint[2][]{arXiv:#2}

\bibitem[Ainsworth(1999)]%
        {ainsworth_functions_1999}
\bibfield{author}{\bibinfo{person}{Shaaron Ainsworth}.} \bibinfo{year}{1999}\natexlab{}.
\newblock \showarticletitle{The functions of multiple representations}.
\newblock \bibinfo{journal}{\emph{Computers \& Education}} \bibinfo{volume}{33}, \bibinfo{number}{2-3} (\bibinfo{date}{Sept.} \bibinfo{year}{1999}), \bibinfo{pages}{131--152}.
\newblock
\showISSN{03601315}
\href{https://doi.org/10.1016/S0360-1315(99)00029-9}{doi:\nolinkurl{10.1016/S0360-1315(99)00029-9}}


\bibitem[Alexander(2003)]%
        {alexander_development_2003}
\bibfield{author}{\bibinfo{person}{Patricia~A. Alexander}.} \bibinfo{year}{2003}\natexlab{}.
\newblock \showarticletitle{The {Development} of {Expertise}: {The} {Journey} {From} {Acclimation} to {Proficiency}}.
\newblock \bibinfo{journal}{\emph{Educational Researcher}} \bibinfo{volume}{32}, \bibinfo{number}{8} (\bibinfo{date}{Nov.} \bibinfo{year}{2003}), \bibinfo{pages}{10--14}.
\newblock
\showISSN{0013-189X, 1935-102X}
\href{https://doi.org/10.3102/0013189X032008010}{doi:\nolinkurl{10.3102/0013189X032008010}}


\bibitem[Alexander(2004)]%
        {alexander_model_2004}
\bibfield{author}{\bibinfo{person}{Patricia~A. Alexander}.} \bibinfo{year}{2004}\natexlab{}.
\newblock \showarticletitle{A {Model} of {Domain} {Learning}: {Reinterpreting} {Expertise} as a {Multidimensional}, {Multistage} {Process}}.
\newblock In \bibinfo{booktitle}{\emph{Motivation, {Emotion}, and {Cognition}}}. \bibinfo{publisher}{Routledge}.
\newblock
\newblock
\shownote{Num Pages: 26}.


\bibitem[Amoozadeh et~al\mbox{.}(2024)]%
        {amoozadeh_trust_2024}
\bibfield{author}{\bibinfo{person}{Matin Amoozadeh}, \bibinfo{person}{David Daniels}, \bibinfo{person}{Daye Nam}, \bibinfo{person}{Aayush Kumar}, \bibinfo{person}{Stella Chen}, \bibinfo{person}{Michael Hilton}, \bibinfo{person}{Sruti Srinivasa~Ragavan}, {and} \bibinfo{person}{Mohammad~Amin Alipour}.} \bibinfo{year}{2024}\natexlab{}.
\newblock \showarticletitle{Trust in {Generative} {AI} among {Students}: {An} exploratory study}. In \bibinfo{booktitle}{\emph{Proceedings of the 55th {ACM} {Technical} {Symposium} on {Computer} {Science} {Education} {V}. 1}} \emph{(\bibinfo{series}{{SIGCSE} 2024})}. \bibinfo{publisher}{Association for Computing Machinery}, \bibinfo{address}{New York, NY, USA}, \bibinfo{pages}{67--73}.
\newblock
\showISBNx{979-8-4007-0423-9}
\href{https://doi.org/10.1145/3626252.3630842}{doi:\nolinkurl{10.1145/3626252.3630842}}


\bibitem[Balse et~al\mbox{.}(2023)]%
        {balse_evaluating_2023}
\bibfield{author}{\bibinfo{person}{Rishabh Balse}, \bibinfo{person}{Viraj Kumar}, \bibinfo{person}{Prajish Prasad}, {and} \bibinfo{person}{Jayakrishnan~Madathil Warriem}.} \bibinfo{year}{2023}\natexlab{}.
\newblock \showarticletitle{Evaluating the {Quality} of {LLM}-{Generated} {Explanations} for {Logical} {Errors} in {CS1} {Student} {Programs}}. In \bibinfo{booktitle}{\emph{Proceedings of the 16th {Annual} {ACM} {India} {Compute} {Conference}}} \emph{(\bibinfo{series}{{COMPUTE} '23})}. \bibinfo{publisher}{Association for Computing Machinery}, \bibinfo{address}{New York, NY, USA}, \bibinfo{pages}{49--54}.
\newblock
\showISBNx{979-8-4007-0840-4}
\href{https://doi.org/10.1145/3627217.3627233}{doi:\nolinkurl{10.1145/3627217.3627233}}


\bibitem[Bernacki and Walkington(2018)]%
        {bernacki_role_2018}
\bibfield{author}{\bibinfo{person}{Matthew~L. Bernacki} {and} \bibinfo{person}{Candace Walkington}.} \bibinfo{year}{2018}\natexlab{}.
\newblock \showarticletitle{The role of situational interest in personalized learning}.
\newblock \bibinfo{journal}{\emph{Journal of Educational Psychology}} \bibinfo{volume}{110}, \bibinfo{number}{6} (\bibinfo{year}{2018}), \bibinfo{pages}{864--881}.
\newblock
\showISSN{1939-2176}
\href{https://doi.org/10.1037/edu0000250}{doi:\nolinkurl{10.1037/edu0000250}}


\bibitem[Bernstein et~al\mbox{.}(2024a)]%
        {bernstein_like_2024}
\bibfield{author}{\bibinfo{person}{Seth Bernstein}, \bibinfo{person}{Paul Denny}, \bibinfo{person}{Juho Leinonen}, \bibinfo{person}{Lauren Kan}, \bibinfo{person}{Arto Hellas}, \bibinfo{person}{Matt Littlefield}, \bibinfo{person}{Sami Sarsa}, {and} \bibinfo{person}{Stephen MacNeil}.} \bibinfo{year}{2024}\natexlab{a}.
\newblock \showarticletitle{"{Like} a {Nesting} {Doll}": {Analyzing} {Recursion} {Analogies} {Generated} by {CS} {Students} {Using} {Large} {Language} {Models}}. In \bibinfo{booktitle}{\emph{Proceedings of the 2024 on {Innovation} and {Technology} in {Computer} {Science} {Education} {V}. 1}} \emph{(\bibinfo{series}{{ITiCSE} 2024})}. \bibinfo{publisher}{Association for Computing Machinery}, \bibinfo{address}{New York, NY, USA}, \bibinfo{pages}{122--128}.
\newblock
\showISBNx{979-8-4007-0600-4}
\href{https://doi.org/10.1145/3649217.3653533}{doi:\nolinkurl{10.1145/3649217.3653533}}


\bibitem[Bernstein et~al\mbox{.}(2024b)]%
        {bernstein_analyzing_2024}
\bibfield{author}{\bibinfo{person}{Seth Bernstein}, \bibinfo{person}{Paul Denny}, \bibinfo{person}{Juho Leinonen}, \bibinfo{person}{Matt Littlefield}, \bibinfo{person}{Arto Hellas}, {and} \bibinfo{person}{Stephen MacNeil}.} \bibinfo{year}{2024}\natexlab{b}.
\newblock \showarticletitle{Analyzing {Students}' {Preferences} for {LLM}-{Generated} {Analogies}}. In \bibinfo{booktitle}{\emph{Proceedings of the 2024 on {Innovation} and {Technology} in {Computer} {Science} {Education} {V}. 2}}. \bibinfo{publisher}{ACM}, \bibinfo{address}{Milan Italy}, \bibinfo{pages}{812--812}.
\newblock
\showISBNx{979-8-4007-0603-5}
\href{https://doi.org/10.1145/3649405.3659504}{doi:\nolinkurl{10.1145/3649405.3659504}}


\bibitem[Bernstein et~al\mbox{.}(2025)]%
        {bernstein_beyond_2025}
\bibfield{author}{\bibinfo{person}{Seth Bernstein}, \bibinfo{person}{Ashfin Rahman}, \bibinfo{person}{Nadia Sharifi}, \bibinfo{person}{Ariunjargal Terbish}, {and} \bibinfo{person}{Stephen MacNeil}.} \bibinfo{year}{2025}\natexlab{}.
\newblock \showarticletitle{Beyond the {Benefits}: {A} {Systematic} {Review} of the {Harms} and {Consequences} of {Generative} {AI} in {Computing} {Education}}. In \bibinfo{booktitle}{\emph{Proceedings of the 25th {Koli} {Calling} {International} {Conference} on {Computing} {Education} {Research}}}. \bibinfo{publisher}{ACM}, \bibinfo{address}{Koli Finland}, \bibinfo{pages}{1--18}.
\newblock
\showISBNx{979-8-4007-1599-0}
\href{https://doi.org/10.1145/3769994.3770036}{doi:\nolinkurl{10.1145/3769994.3770036}}


\bibitem[Bettin and Ott(2023)]%
        {bettin_pedagogical_2023}
\bibfield{author}{\bibinfo{person}{Briana Bettin} {and} \bibinfo{person}{Linda Ott}.} \bibinfo{year}{2023}\natexlab{}.
\newblock \showarticletitle{Pedagogical {Prisms}: {Toward} {Domain} {Isomorphic} {Analogy} {Design} for {Relevance} and {Engagement} in {Computing} {Education}}. In \bibinfo{booktitle}{\emph{Proceedings of the 2023 {Conference} on {Innovation} and {Technology} in {Computer} {Science} {Education} {V}. 1}} \emph{(\bibinfo{series}{{ITiCSE} 2023})}. \bibinfo{publisher}{Association for Computing Machinery}, \bibinfo{address}{New York, NY, USA}, \bibinfo{pages}{410--416}.
\newblock
\showISBNx{979-8-4007-0138-2}
\href{https://doi.org/10.1145/3587102.3588830}{doi:\nolinkurl{10.1145/3587102.3588830}}


\bibitem[Bettin et~al\mbox{.}(2022)]%
        {bettin2022semaphore}
\bibfield{author}{\bibinfo{person}{Briana Bettin}, \bibinfo{person}{Linda Ott}, {and} \bibinfo{person}{Julia Hiebel}.} \bibinfo{year}{2022}\natexlab{}.
\newblock \showarticletitle{Semaphore or Metaphor? Exploring Concurrent Students' Conceptions of and With Analogy}. In \bibinfo{booktitle}{\emph{Proceedings of the 27th ACM Conference on on Innovation and Technology in Computer Science Education Vol. 1}} (Dublin, Ireland) \emph{(\bibinfo{series}{ITiCSE '22})}. \bibinfo{publisher}{Association for Computing Machinery}, \bibinfo{address}{New York, NY, USA}, \bibinfo{pages}{200–206}.
\newblock
\showISBNx{9781450392013}
\href{https://doi.org/10.1145/3502718.3524796}{doi:\nolinkurl{10.1145/3502718.3524796}}


\bibitem[Bettin et~al\mbox{.}(2023)]%
        {bettin2023more}
\bibfield{author}{\bibinfo{person}{Briana Bettin}, \bibinfo{person}{Linda Ott}, {and} \bibinfo{person}{Julia Hiebel}.} \bibinfo{year}{2023}\natexlab{}.
\newblock \showarticletitle{More (Sema|Meta)phors: Additional Perspectives on Analogy Use From Concurrent Programming Students}. In \bibinfo{booktitle}{\emph{Proceedings of the 2023 Conference on Innovation and Technology in Computer Science Education v. 1}} (Turku, Finland) \emph{(\bibinfo{series}{ITiCSE 2023})}. \bibinfo{publisher}{Association for Computing Machinery}, \bibinfo{address}{New York, NY, USA}, \bibinfo{pages}{166–172}.
\newblock
\showISBNx{9798400701382}
\href{https://doi.org/10.1145/3587102.3588831}{doi:\nolinkurl{10.1145/3587102.3588831}}


\bibitem[Boustedt et~al\mbox{.}(2007)]%
        {boustedt2007threshold}
\bibfield{author}{\bibinfo{person}{Jonas Boustedt}, \bibinfo{person}{Anna Eckerdal}, \bibinfo{person}{Robert McCartney}, \bibinfo{person}{Jan~Erik Mostr\"{o}m}, \bibinfo{person}{Mark Ratcliffe}, \bibinfo{person}{Kate Sanders}, {and} \bibinfo{person}{Carol Zander}.} \bibinfo{year}{2007}\natexlab{}.
\newblock \showarticletitle{Threshold Concepts in Computer Science: Do They Exist and Are They Useful?}. In \bibinfo{booktitle}{\emph{Proceedings of the 38th SIGCSE Technical Symposium on Computer Science Education}} (Covington, Kentucky, USA) \emph{(\bibinfo{series}{SIGCSE '07})}. \bibinfo{publisher}{Association for Computing Machinery}, \bibinfo{address}{New York, NY, USA}, \bibinfo{pages}{504–508}.
\newblock
\showISBNx{1595933611}
\href{https://doi.org/10.1145/1227310.1227482}{doi:\nolinkurl{10.1145/1227310.1227482}}


\bibitem[Braun and Clarke(2006)]%
        {braun_using_2006}
\bibfield{author}{\bibinfo{person}{Virginia Braun} {and} \bibinfo{person}{Victoria Clarke}.} \bibinfo{year}{2006}\natexlab{}.
\newblock \showarticletitle{Using thematic analysis in psychology}.
\newblock \bibinfo{journal}{\emph{Qualitative Research in Psychology}} \bibinfo{volume}{3}, \bibinfo{number}{2} (\bibinfo{date}{Jan.} \bibinfo{year}{2006}), \bibinfo{pages}{77--101}.
\newblock
\showISSN{1478-0887, 1478-0895}
\href{https://doi.org/10.1191/1478088706qp063oa}{doi:\nolinkurl{10.1191/1478088706qp063oa}}


\bibitem[Braun et~al\mbox{.}(2023)]%
        {braun2023doing}
\bibfield{author}{\bibinfo{person}{Virginia Braun}, \bibinfo{person}{Victoria Clarke}, \bibinfo{person}{Nikki Hayfield}, \bibinfo{person}{Louise Davey}, {and} \bibinfo{person}{Elizabeth Jenkinson}.} \bibinfo{year}{2023}\natexlab{}.
\newblock \showarticletitle{Doing Reflexive Thematic Analysis}.
\newblock In \bibinfo{booktitle}{\emph{Supporting Research in Counselling and Psychotherapy: Qualitative, Quantitative, and Mixed Methods Research}}. \bibinfo{publisher}{Springer}, \bibinfo{address}{Berlin, Germany}, \bibinfo{pages}{19--38}.
\newblock
\href{https://doi.org/10.1007/978-3-031-13942-0_2}{doi:\nolinkurl{10.1007/978-3-031-13942-0_2}}


\bibitem[Cao et~al\mbox{.}(2016)]%
        {cao2016examining}
\bibfield{author}{\bibinfo{person}{Yingjun Cao}, \bibinfo{person}{Leo Porter}, {and} \bibinfo{person}{Daniel Zingaro}.} \bibinfo{year}{2016}\natexlab{}.
\newblock \showarticletitle{Examining the Value of Analogies in Introductory Computing}. In \bibinfo{booktitle}{\emph{Proceedings of the 2016 ACM Conference on International Computing Education Research}}. \bibinfo{publisher}{ACM}, \bibinfo{address}{New York, NY, USA}, \bibinfo{pages}{231--239}.
\newblock
\href{https://doi.org/10.1145/2960310.2960313}{doi:\nolinkurl{10.1145/2960310.2960313}}


\bibitem[Cardell-Oliver et~al\mbox{.}(2026)]%
        {cardell2026llm}
\bibfield{author}{\bibinfo{person}{Rachel Cardell-Oliver}, \bibinfo{person}{Claudia Szabo}, \bibinfo{person}{Kaie Maennel}, {and} \bibinfo{person}{Hamish Russell}.} \bibinfo{year}{2026}\natexlab{}.
\newblock \showarticletitle{LLM-led Socratic Dialogues on Ethics in Computing}. In \bibinfo{booktitle}{\emph{Proceedings of the 31st ACM Conference on Innovation and Technology in Computer Science Education v. 1}}. \bibinfo{publisher}{ACM}, \bibinfo{address}{New York, NY, USA}, \bibinfo{pages}{121--127}.
\newblock
\href{https://doi.org/10.1145/3803400.3809325}{doi:\nolinkurl{10.1145/3803400.3809325}}


\bibitem[Clancy(2005)]%
        {clancy2005misconceptions}
\bibfield{author}{\bibinfo{person}{Michael Clancy}.} \bibinfo{year}{2005}\natexlab{}.
\newblock \showarticletitle{Misconceptions and Attitudes That Interfere With Learning to Program}.
\newblock In \bibinfo{booktitle}{\emph{Computer Science Education Research}}. \bibinfo{publisher}{Taylor \& Francis}, \bibinfo{address}{London, UK}, \bibinfo{pages}{95--110}.
\newblock
\href{https://doi.org/10.1201/9781482287325-18}{doi:\nolinkurl{10.1201/9781482287325-18}}


\bibitem[Del Carpio~Gutierrez et~al\mbox{.}(2024)]%
        {del_carpio_gutierrez_evaluating_2024}
\bibfield{author}{\bibinfo{person}{Andre Del Carpio~Gutierrez}, \bibinfo{person}{Paul Denny}, {and} \bibinfo{person}{Andrew Luxton-Reilly}.} \bibinfo{year}{2024}\natexlab{}.
\newblock \showarticletitle{Evaluating {Automatically} {Generated} {Contextualised} {Programming} {Exercises}}. In \bibinfo{booktitle}{\emph{Proceedings of the 55th {ACM} {Technical} {Symposium} on {Computer} {Science} {Education} {V}. 1}} \emph{(\bibinfo{series}{{SIGCSE} 2024})}. \bibinfo{publisher}{Association for Computing Machinery}, \bibinfo{address}{New York, NY, USA}, \bibinfo{pages}{289--295}.
\newblock
\showISBNx{979-8-4007-0423-9}
\href{https://doi.org/10.1145/3626252.3630863}{doi:\nolinkurl{10.1145/3626252.3630863}}


\bibitem[Ericson et~al\mbox{.}(2019)]%
        {ericson_investigating_2019}
\bibfield{author}{\bibinfo{person}{Barbara Ericson}, \bibinfo{person}{Austin McCall}, {and} \bibinfo{person}{Kathryn Cunningham}.} \bibinfo{year}{2019}\natexlab{}.
\newblock \showarticletitle{Investigating the {Affect} and {Effect} of {Adaptive} {Parsons} {Problems}}. In \bibinfo{booktitle}{\emph{Proceedings of the 19th {Koli} {Calling} {International} {Conference} on {Computing} {Education} {Research}}} \emph{(\bibinfo{series}{Koli {Calling} '19})}. \bibinfo{publisher}{Association for Computing Machinery}, \bibinfo{address}{New York, NY, USA}, \bibinfo{pages}{1--10}.
\newblock
\showISBNx{978-1-4503-7715-7}
\href{https://doi.org/10.1145/3364510.3364524}{doi:\nolinkurl{10.1145/3364510.3364524}}


\bibitem[Gentner(1983)]%
        {gentner_structure-mapping_1983}
\bibfield{author}{\bibinfo{person}{Dedre Gentner}.} \bibinfo{year}{1983}\natexlab{}.
\newblock \showarticletitle{Structure-mapping: {A} theoretical framework for analogy}.
\newblock \bibinfo{journal}{\emph{Cognitive Science}} \bibinfo{volume}{7}, \bibinfo{number}{2} (\bibinfo{date}{April} \bibinfo{year}{1983}), \bibinfo{pages}{155--170}.
\newblock
\showISSN{0364-0213}
\href{https://doi.org/10.1016/S0364-0213(83)80009-3}{doi:\nolinkurl{10.1016/S0364-0213(83)80009-3}}


\bibitem[Good et~al\mbox{.}(2012)]%
        {good2012belonging}
\bibfield{author}{\bibinfo{person}{Catherine Good}, \bibinfo{person}{Aneeta Rattan}, {and} \bibinfo{person}{Carol~S Dweck}.} \bibinfo{year}{2012}\natexlab{}.
\newblock \showarticletitle{Why Do Women Opt Out? Sense of Belonging and Women's Representation in Mathematics.}
\newblock \bibinfo{journal}{\emph{Journal of Personality and Social Psychology}} \bibinfo{volume}{102}, \bibinfo{number}{4} (\bibinfo{year}{2012}), \bibinfo{pages}{700}.
\newblock
\href{https://doi.org/10.1037/a0026659}{doi:\nolinkurl{10.1037/a0026659}}


\bibitem[Guzdial(2013)]%
        {guzdial_exploring_2013}
\bibfield{author}{\bibinfo{person}{Mark Guzdial}.} \bibinfo{year}{2013}\natexlab{}.
\newblock \showarticletitle{Exploring hypotheses about media computation}. In \bibinfo{booktitle}{\emph{Proceedings of the ninth annual international {ACM} conference on {International} computing education research}} \emph{(\bibinfo{series}{{ICER} '13})}. \bibinfo{publisher}{Association for Computing Machinery}, \bibinfo{address}{New York, NY, USA}, \bibinfo{pages}{19--26}.
\newblock
\showISBNx{978-1-4503-2243-0}
\href{https://doi.org/10.1145/2493394.2493397}{doi:\nolinkurl{10.1145/2493394.2493397}}


\bibitem[Guzdial and Nelson-Fromm(2025)]%
        {guzdial_designing_2025}
\bibfield{author}{\bibinfo{person}{Mark Guzdial} {and} \bibinfo{person}{Tamara Nelson-Fromm}.} \bibinfo{year}{2025}\natexlab{}.
\newblock \showarticletitle{Designing {Courses} for {Liberal} {Arts} and {Sciences} {Students} {Contextualized} around {Creative} {Expression} and {Social} {Justice}}. In \bibinfo{booktitle}{\emph{Proceedings of the 56th {ACM} {Technical} {Symposium} on {Computer} {Science} {Education} {V}. 1}} \emph{(\bibinfo{series}{{SIGCSETS} 2025})}. \bibinfo{publisher}{Association for Computing Machinery}, \bibinfo{address}{New York, NY, USA}, \bibinfo{pages}{423--429}.
\newblock
\showISBNx{979-8-4007-0531-1}
\href{https://doi.org/10.1145/3641554.3701896}{doi:\nolinkurl{10.1145/3641554.3701896}}


\bibitem[Halasz and Moran(1982)]%
        {halasz1982analogy}
\bibfield{author}{\bibinfo{person}{Frank Halasz} {and} \bibinfo{person}{Thomas~P Moran}.} \bibinfo{year}{1982}\natexlab{}.
\newblock \showarticletitle{Analogy Considered Harmful}. In \bibinfo{booktitle}{\emph{Proceedings of the 1982 Conference on Human Factors in Computing Systems}}. \bibinfo{publisher}{ACM Press}, \bibinfo{address}{New York, NY, USA}, \bibinfo{pages}{383--386}.
\newblock
\href{https://doi.org/10.1145/800049.801816}{doi:\nolinkurl{10.1145/800049.801816}}


\bibitem[Hallal et~al\mbox{.}(2026)]%
        {hallal2026beyond}
\bibfield{author}{\bibinfo{person}{Kassem Hallal}, \bibinfo{person}{Rasha Hamdan}, {and} \bibinfo{person}{Sami Tlais}.} \bibinfo{year}{2026}\natexlab{}.
\newblock \showarticletitle{Beyond Traditional Texts: Exploring AI as a Tool for Generating Stories and Analogies in Organic Chemistry Education}.
\newblock \bibinfo{journal}{\emph{Journal of Chemical Education}} \bibinfo{volume}{103}, \bibinfo{number}{2} (\bibinfo{year}{2026}), \bibinfo{pages}{846--857}.
\newblock
\href{https://doi.org/10.1021/acs.jchemed.5c00693}{doi:\nolinkurl{10.1021/acs.jchemed.5c00693}}


\bibitem[Harper et~al\mbox{.}(2023)]%
        {harper2023investigating}
\bibfield{author}{\bibinfo{person}{Colton Harper}, \bibinfo{person}{Ryan Bockmon}, {and} \bibinfo{person}{Stephen Cooper}.} \bibinfo{year}{2023}\natexlab{}.
\newblock \showarticletitle{Investigating Themes of Student-Generated Analogies}. In \bibinfo{booktitle}{\emph{Proceedings of the ACM Conference on Global Computing Education Vol 1}} (Hyderabad, India) \emph{(\bibinfo{series}{CompEd 2023})}. \bibinfo{publisher}{Association for Computing Machinery}, \bibinfo{address}{New York, NY, USA}, \bibinfo{pages}{64–70}.
\newblock
\showISBNx{9798400700484}
\href{https://doi.org/10.1145/3576882.3617914}{doi:\nolinkurl{10.1145/3576882.3617914}}


\bibitem[Heinonen et~al\mbox{.}(2023)]%
        {heinonen2023synthesizing}
\bibfield{author}{\bibinfo{person}{Ava Heinonen}, \bibinfo{person}{Bettina Lehtel{\"a}}, \bibinfo{person}{Arto Hellas}, {and} \bibinfo{person}{Fabian Fagerholm}.} \bibinfo{year}{2023}\natexlab{}.
\newblock \showarticletitle{Synthesizing Research on Programmers’ Mental Models of Programs, Tasks and Concepts—A Systematic Literature Review}.
\newblock \bibinfo{journal}{\emph{Information and Software Technology}}  \bibinfo{volume}{164} (\bibinfo{year}{2023}), \bibinfo{pages}{107300}.
\newblock
\href{https://doi.org/10.1016/j.infsof.2023.107300}{doi:\nolinkurl{10.1016/j.infsof.2023.107300}}


\bibitem[Hidi and Renninger(2006)]%
        {hidi_four-phase_2006}
\bibfield{author}{\bibinfo{person}{Suzanne Hidi} {and} \bibinfo{person}{K.~Ann Renninger}.} \bibinfo{year}{2006}\natexlab{}.
\newblock \showarticletitle{The {Four}-{Phase} {Model} of {Interest} {Development}}.
\newblock \bibinfo{journal}{\emph{Educational Psychologist}} \bibinfo{volume}{41}, \bibinfo{number}{2} (\bibinfo{date}{June} \bibinfo{year}{2006}), \bibinfo{pages}{111--127}.
\newblock
\showISSN{0046-1520}
\href{https://doi.org/10.1207/s15326985ep4102_4}{doi:\nolinkurl{10.1207/s15326985ep4102_4}}
\newblock
\shownote{\_eprint: https://doi.org/10.1207/s15326985ep4102\_4}.


\bibitem[Hou et~al\mbox{.}(2025)]%
        {hou_personalized_2025}
\bibfield{author}{\bibinfo{person}{Xinying Hou}, \bibinfo{person}{Zihan Wu}, \bibinfo{person}{Xu Wang}, {and} \bibinfo{person}{Barbara~J. Ericson}.} \bibinfo{year}{2025}\natexlab{}.
\newblock \showarticletitle{Personalized {Parsons} {Puzzles} as {Scaffolding} {Enhance} {Practice} {Engagement} {Over} {Just} {Showing} {LLM}-{Powered} {Solutions}}. In \bibinfo{booktitle}{\emph{Proceedings of the 56th {ACM} {Technical} {Symposium} on {Computer} {Science} {Education} {V}. 2}} \emph{(\bibinfo{series}{{SIGCSETS} 2025})}. \bibinfo{publisher}{Association for Computing Machinery}, \bibinfo{address}{New York, NY, USA}, \bibinfo{pages}{1483--1484}.
\newblock
\showISBNx{979-8-4007-0532-8}
\href{https://doi.org/10.1145/3641555.3705227}{doi:\nolinkurl{10.1145/3641555.3705227}}


\bibitem[Krakauer(2016)]%
        {krakauer2016will}
\bibfield{author}{\bibinfo{person}{David Krakauer}.} \bibinfo{year}{2016}\natexlab{}.
\newblock \showarticletitle{Will {AI} Harm Us? {B}etter to Ask How We'll Reckon With Our Hybrid Nature}.
\newblock \bibinfo{journal}{\emph{Nautilus}}  \bibinfo{volume}{6} (\bibinfo{year}{2016}).
\newblock


\bibitem[Kunda(1990)]%
        {kunda1990case}
\bibfield{author}{\bibinfo{person}{Ziva Kunda}.} \bibinfo{year}{1990}\natexlab{}.
\newblock \showarticletitle{The Case for Motivated Reasoning}.
\newblock \bibinfo{journal}{\emph{Psychological Bulletin}} \bibinfo{volume}{108}, \bibinfo{number}{3} (\bibinfo{year}{1990}), \bibinfo{pages}{480--498}.
\newblock
\showISSN{1939-1455}
\href{https://doi.org/10.1037/0033-2909.108.3.480}{doi:\nolinkurl{10.1037/0033-2909.108.3.480}}


\bibitem[Layman et~al\mbox{.}(2020)]%
        {layman2020toward}
\bibfield{author}{\bibinfo{person}{Lucas Layman}, \bibinfo{person}{Yang Song}, {and} \bibinfo{person}{Curry Guinn}.} \bibinfo{year}{2020}\natexlab{}.
\newblock \showarticletitle{Toward Predicting Success and Failure in CS2: A Mixed-Method Analysis}. In \bibinfo{booktitle}{\emph{Proceedings of the 2020 ACM Southeast Conference}} (Tampa, FL, USA) \emph{(\bibinfo{series}{ACMSE '20})}. \bibinfo{publisher}{Association for Computing Machinery}, \bibinfo{address}{New York, NY, USA}, \bibinfo{pages}{218–225}.
\newblock
\showISBNx{9781450371056}
\href{https://doi.org/10.1145/3374135.3385277}{doi:\nolinkurl{10.1145/3374135.3385277}}


\bibitem[Lee and Song(2024)]%
        {lee_teachers_2024}
\bibfield{author}{\bibinfo{person}{Soohwan Lee} {and} \bibinfo{person}{Ki-Sang Song}.} \bibinfo{year}{2024}\natexlab{}.
\newblock \showarticletitle{Teachers' and students' perceptions of {AI}-generated concept explanations: {Implications} for integrating generative {AI} in computer science education}.
\newblock \bibinfo{journal}{\emph{Computers and Education: Artificial Intelligence}}  \bibinfo{volume}{7} (\bibinfo{year}{2024}).
\newblock
\href{https://doi.org/10.1016/j.caeai.2024.100283}{doi:\nolinkurl{10.1016/j.caeai.2024.100283}}
\newblock
\shownote{Type: Article}.


\bibitem[Leinonen et~al\mbox{.}(2023)]%
        {leinonen_comparing_2023}
\bibfield{author}{\bibinfo{person}{Juho Leinonen}, \bibinfo{person}{Paul Denny}, \bibinfo{person}{Stephen MacNeil}, \bibinfo{person}{Sami Sarsa}, \bibinfo{person}{Seth Bernstein}, \bibinfo{person}{Joanne Kim}, \bibinfo{person}{Andrew Tran}, {and} \bibinfo{person}{Arto Hellas}.} \bibinfo{year}{2023}\natexlab{}.
\newblock \bibinfo{title}{Comparing {Code} {Explanations} {Created} by {Students} and {Large} {Language} {Models}}.
\newblock
\href{https://doi.org/10.1145/3587102.3588785}{doi:\nolinkurl{10.1145/3587102.3588785}}
\newblock
\shownote{Pages: 124–130 Publication Title: Proceedings of the 2023 Conference on Innovation and Technology in Computer Science Education V. 1}.


\bibitem[Lewis et~al\mbox{.}(2017)]%
        {lewis2017fitting}
\bibfield{author}{\bibinfo{person}{Karyn~L Lewis}, \bibinfo{person}{Jane~G Stout}, \bibinfo{person}{Noah~D Finkelstein}, \bibinfo{person}{Steven~J Pollock}, \bibinfo{person}{Akira Miyake}, \bibinfo{person}{Geoff~L Cohen}, {and} \bibinfo{person}{Tiffany~A Ito}.} \bibinfo{year}{2017}\natexlab{}.
\newblock \showarticletitle{Fitting in to Move Forward: Belonging, Gender, and Persistence in the Physical Sciences, Technology, Engineering, and Mathematics (pSTEM)}.
\newblock \bibinfo{journal}{\emph{Psychology of Women Quarterly}} \bibinfo{volume}{41}, \bibinfo{number}{4} (\bibinfo{year}{2017}), \bibinfo{pages}{420--436}.
\newblock
\href{https://doi.org/10.1177/0361684317720186}{doi:\nolinkurl{10.1177/0361684317720186}}


\bibitem[Liu et~al\mbox{.}(2026)]%
        {liu_analogical_2026}
\bibfield{author}{\bibinfo{person}{Jonathan Liu}, \bibinfo{person}{Erica Goodwin}, {and} \bibinfo{person}{Diana Franklin}.} \bibinfo{year}{2026}\natexlab{}.
\newblock \showarticletitle{Analogical {Reasoning} in {Undergraduate} {Algorithms}}.
\newblock In \bibinfo{booktitle}{\emph{Proceedings of the 57th {ACM} {Technical} {Symposium} on {Computer} {Science} {Education} {V}.1}}. Vol.~\bibinfo{volume}{1}. \bibinfo{publisher}{Association for Computing Machinery}, \bibinfo{address}{New York, NY, USA}, \bibinfo{pages}{673--679}.
\newblock
\showISBNx{979-8-4007-2256-1}
\urldef\tempurl%
\url{https://dl.acm.org/doi/10.1145/3770762.3772624}
\showURL{%
\tempurl}


\bibitem[Liu and M'Hiri(2024)]%
        {liu_beyond_2024}
\bibfield{author}{\bibinfo{person}{Mengqi Liu} {and} \bibinfo{person}{Faten M'Hiri}.} \bibinfo{year}{2024}\natexlab{}.
\newblock \showarticletitle{Beyond {Traditional} {Teaching}: {Large} {Language} {Models} as {Simulated} {Teaching} {Assistants} in {Computer} {Science}}. In \bibinfo{booktitle}{\emph{Proceedings of the 55th {ACM} {Technical} {Symposium} on {Computer} {Science} {Education} {V}. 1}} \emph{(\bibinfo{series}{{SIGCSE} 2024})}. \bibinfo{publisher}{Association for Computing Machinery}, \bibinfo{address}{New York, NY, USA}, \bibinfo{pages}{743--749}.
\newblock
\showISBNx{979-8-4007-0423-9}
\href{https://doi.org/10.1145/3626252.3630789}{doi:\nolinkurl{10.1145/3626252.3630789}}


\bibitem[Logacheva et~al\mbox{.}(2024)]%
        {logacheva_evaluating_2024}
\bibfield{author}{\bibinfo{person}{Evanfiya Logacheva}, \bibinfo{person}{Arto Hellas}, \bibinfo{person}{James Prather}, \bibinfo{person}{Sami Sarsa}, {and} \bibinfo{person}{Juho Leinonen}.} \bibinfo{year}{2024}\natexlab{}.
\newblock \bibinfo{title}{Evaluating {Contextually} {Personalized} {Programming} {Exercises} {Created} with {Generative} {AI}}.
\newblock
\href{https://doi.org/10.1145/3632620.3671103}{doi:\nolinkurl{10.1145/3632620.3671103}}
\newblock
\shownote{Pages: 95–113 Volume: 1}.


\bibitem[MacNeil et~al\mbox{.}(2023)]%
        {macneil_experiences_2023}
\bibfield{author}{\bibinfo{person}{Stephen MacNeil}, \bibinfo{person}{Andrew Tran}, \bibinfo{person}{Arto Hellas}, \bibinfo{person}{Joanne Kim}, \bibinfo{person}{Sami Sarsa}, \bibinfo{person}{Paul Denny}, \bibinfo{person}{Seth Bernstein}, {and} \bibinfo{person}{Juho Leinonen}.} \bibinfo{year}{2023}\natexlab{}.
\newblock \showarticletitle{Experiences from {Using} {Code} {Explanations} {Generated} by {Large} {Language} {Models} in a {Web} {Software} {Development} {E}-{Book}}. In \bibinfo{booktitle}{\emph{Proceedings of the 54th {ACM} {Technical} {Symposium} on {Computer} {Science} {Education} {V}. 1}} \emph{(\bibinfo{series}{{SIGCSE} 2023})}. \bibinfo{publisher}{Association for Computing Machinery}, \bibinfo{address}{New York, NY, USA}, \bibinfo{pages}{931--937}.
\newblock
\showISBNx{978-1-4503-9431-4}
\href{https://doi.org/10.1145/3545945.3569785}{doi:\nolinkurl{10.1145/3545945.3569785}}


\bibitem[MacNeil et~al\mbox{.}(2022)]%
        {macneil_generating_2022}
\bibfield{author}{\bibinfo{person}{Stephen MacNeil}, \bibinfo{person}{Andrew Tran}, \bibinfo{person}{Dan Mogil}, \bibinfo{person}{Seth Bernstein}, \bibinfo{person}{Erin Ross}, {and} \bibinfo{person}{Ziheng Huang}.} \bibinfo{year}{2022}\natexlab{}.
\newblock \showarticletitle{Generating {Diverse} {Code} {Explanations} using the {GPT}-3 {Large} {Language} {Model}}. In \bibinfo{booktitle}{\emph{Proceedings of the 2022 {ACM} {Conference} on {International} {Computing} {Education} {Research} - {Volume} 2}} \emph{(\bibinfo{series}{{ICER} '22}, Vol.~\bibinfo{volume}{2})}. \bibinfo{publisher}{Association for Computing Machinery}, \bibinfo{address}{New York, NY, USA}, \bibinfo{pages}{37--39}.
\newblock
\showISBNx{978-1-4503-9195-5}
\href{https://doi.org/10.1145/3501709.3544280}{doi:\nolinkurl{10.1145/3501709.3544280}}


\bibitem[Master et~al\mbox{.}(2016)]%
        {master2016computing}
\bibfield{author}{\bibinfo{person}{Allison Master}, \bibinfo{person}{Sapna Cheryan}, {and} \bibinfo{person}{Andrew~N Meltzoff}.} \bibinfo{year}{2016}\natexlab{}.
\newblock \showarticletitle{Computing Whether She Belongs: Stereotypes Undermine Girls’ Interest and Sense of Belonging in Computer Science.}
\newblock \bibinfo{journal}{\emph{Journal of Educational Psychology}} \bibinfo{volume}{108}, \bibinfo{number}{3} (\bibinfo{year}{2016}), \bibinfo{pages}{424}.
\newblock
\href{https://doi.org/10.1037/edu0000061}{doi:\nolinkurl{10.1037/edu0000061}}


\bibitem[Padiyath and Nelson-Fromm(2026)]%
        {padiyath2026reflecting}
\bibfield{author}{\bibinfo{person}{Aadarsh Padiyath} {and} \bibinfo{person}{Tamara Nelson-Fromm}.} \bibinfo{year}{2026}\natexlab{}.
\newblock \showarticletitle{Reflecting on Thematic Analysis in Computer Science Education Research: A Field Guide for Researchers and Reviewers}. In \bibinfo{booktitle}{\emph{Proceedings of the 57th ACM Technical Symposium on Computer Science Education V.1}} (USA) \emph{(\bibinfo{series}{SIGCSE TS 2026})}. \bibinfo{publisher}{Association for Computing Machinery}, \bibinfo{address}{New York, NY, USA}, \bibinfo{pages}{790–796}.
\newblock
\showISBNx{9798400722561}
\href{https://doi.org/10.1145/3770762.3772512}{doi:\nolinkurl{10.1145/3770762.3772512}}


\bibitem[Paul and Elder(2006)]%
        {paul2006critical}
\bibfield{author}{\bibinfo{person}{Richard Paul} {and} \bibinfo{person}{Linda Elder}.} \bibinfo{year}{2006}\natexlab{}.
\newblock \showarticletitle{Critical Thinking: The Nature of Critical and Creative Thought}.
\newblock \bibinfo{journal}{\emph{Journal of Developmental Education}} \bibinfo{volume}{30}, \bibinfo{number}{2} (\bibinfo{year}{2006}), \bibinfo{pages}{34}.
\newblock


\bibitem[Piaget(1970)]%
        {piaget1970science}
\bibfield{author}{\bibinfo{person}{Jean Piaget}.} \bibinfo{year}{1970}\natexlab{}.
\newblock \showarticletitle{Science of Education and the Psychology of the Child. Trans. D. Coltman.}
\newblock  (\bibinfo{year}{1970}).
\newblock


\bibitem[Pitts et~al\mbox{.}(2026)]%
        {pitts_personalized_2026}
\bibfield{author}{\bibinfo{person}{Griffin Pitts}, \bibinfo{person}{Muntasir Hoq}, \bibinfo{person}{Peter Brusilovsky}, \bibinfo{person}{Narges Norouzi}, \bibinfo{person}{Arto Hellas}, \bibinfo{person}{Juho Leinonen}, {and} \bibinfo{person}{Bita Akram}.} \bibinfo{year}{2026}\natexlab{}.
\newblock \bibinfo{title}{Personalized {Worked} {Example} {Generation} from {Student} {Code} {Submissions} using {Pattern}-based {Knowledge} {Components}}.
\newblock
\href{https://doi.org/10.48550/arXiv.2604.24758}{doi:\nolinkurl{10.48550/arXiv.2604.24758}}
\newblock
\shownote{arXiv:2604.24758 [cs]}.


\bibitem[Prather et~al\mbox{.}(2025)]%
        {prather_beyond_2025}
\bibfield{author}{\bibinfo{person}{James Prather}, \bibinfo{person}{Juho Leinonen}, \bibinfo{person}{Natalie Kiesler}, \bibinfo{person}{Jamie Gorson~Benario}, \bibinfo{person}{Sam Lau}, \bibinfo{person}{Stephen MacNeil}, \bibinfo{person}{Narges Norouzi}, \bibinfo{person}{Simone Opel}, \bibinfo{person}{Vee Pettit}, \bibinfo{person}{Leo Porter}, \bibinfo{person}{Brent~N. Reeves}, \bibinfo{person}{Jaromir Savelka}, \bibinfo{person}{David~H. Smith}, \bibinfo{person}{Sven Strickroth}, {and} \bibinfo{person}{Daniel Zingaro}.} \bibinfo{year}{2025}\natexlab{}.
\newblock \showarticletitle{Beyond the {Hype}: {A} {Comprehensive} {Review} of {Current} {Trends} in {Generative} {AI} {Research}, {Teaching} {Practices}, and {Tools}}. In \bibinfo{booktitle}{\emph{2024 {Working} {Group} {Reports} on {Innovation} and {Technology} in {Computer} {Science} {Education}}} \emph{(\bibinfo{series}{{ITiCSE} 2024})}. \bibinfo{publisher}{Association for Computing Machinery}, \bibinfo{address}{New York, NY, USA}, \bibinfo{pages}{300--338}.
\newblock
\showISBNx{979-8-4007-1208-1}
\href{https://doi.org/10.1145/3689187.3709614}{doi:\nolinkurl{10.1145/3689187.3709614}}


\bibitem[Prather et~al\mbox{.}(2024)]%
        {prather_widening_2024}
\bibfield{author}{\bibinfo{person}{James Prather}, \bibinfo{person}{Brent~N Reeves}, \bibinfo{person}{Juho Leinonen}, \bibinfo{person}{Stephen MacNeil}, \bibinfo{person}{Arisoa~S Randrianasolo}, \bibinfo{person}{Brett~A. Becker}, \bibinfo{person}{Bailey Kimmel}, \bibinfo{person}{Jared Wright}, {and} \bibinfo{person}{Ben Briggs}.} \bibinfo{year}{2024}\natexlab{}.
\newblock \bibinfo{title}{The {Widening} {Gap}: {The} {Benefits} and {Harms} of {Generative} {AI} for {Novice} {Programmers}}.
\newblock
\href{https://doi.org/10.1145/3632620.3671116}{doi:\nolinkurl{10.1145/3632620.3671116}}
\newblock
\shownote{Pages: 469–486 Volume: 1}.


\bibitem[Sanders and McCartney(2016)]%
        {sanders2016threshold}
\bibfield{author}{\bibinfo{person}{Kate Sanders} {and} \bibinfo{person}{Robert McCartney}.} \bibinfo{year}{2016}\natexlab{}.
\newblock \showarticletitle{Threshold Concepts in Computing: Past, Present, and Future}. In \bibinfo{booktitle}{\emph{Proceedings of the 16th Koli Calling International Conference on Computing Education Research}} (Koli, Finland) \emph{(\bibinfo{series}{Koli Calling '16})}. \bibinfo{publisher}{Association for Computing Machinery}, \bibinfo{address}{New York, NY, USA}, \bibinfo{pages}{91–100}.
\newblock
\showISBNx{9781450347709}
\href{https://doi.org/10.1145/2999541.2999546}{doi:\nolinkurl{10.1145/2999541.2999546}}


\bibitem[Sarsa et~al\mbox{.}(2022)]%
        {sarsa_automatic_2022}
\bibfield{author}{\bibinfo{person}{Sami Sarsa}, \bibinfo{person}{Paul Denny}, \bibinfo{person}{Arto Hellas}, {and} \bibinfo{person}{Juho Leinonen}.} \bibinfo{year}{2022}\natexlab{}.
\newblock \showarticletitle{Automatic {Generation} of {Programming} {Exercises} and {Code} {Explanations} {Using} {Large} {Language} {Models}}. In \bibinfo{booktitle}{\emph{Proceedings of the 2022 {ACM} {Conference} on {International} {Computing} {Education} {Research} - {Volume} 1}} \emph{(\bibinfo{series}{{ICER} '22})}. \bibinfo{publisher}{Association for Computing Machinery}, \bibinfo{address}{New York, NY, USA}, \bibinfo{pages}{27--43}.
\newblock
\showISBNx{978-1-4503-9194-8}
\href{https://doi.org/10.1145/3501385.3543957}{doi:\nolinkurl{10.1145/3501385.3543957}}


\bibitem[Smith and Zilles(2024)]%
        {smith_evaluating_2024}
\bibfield{author}{\bibinfo{person}{David~H. Smith} {and} \bibinfo{person}{Craig Zilles}.} \bibinfo{year}{2024}\natexlab{}.
\newblock \showarticletitle{Evaluating {Large} {Language} {Model} {Code} {Generation} as an {Autograding} {Mechanism} for "{Explain} in {Plain} {English}" {Questions}}. In \bibinfo{booktitle}{\emph{Proceedings of the 55th {ACM} {Technical} {Symposium} on {Computer} {Science} {Education} {V}. 2}} \emph{(\bibinfo{series}{{SIGCSE} 2024})}. \bibinfo{publisher}{Association for Computing Machinery}, \bibinfo{address}{New York, NY, USA}, \bibinfo{pages}{1824--1825}.
\newblock
\showISBNx{979-8-4007-0424-6}
\href{https://doi.org/10.1145/3626253.3635542}{doi:\nolinkurl{10.1145/3626253.3635542}}


\bibitem[Sorva et~al\mbox{.}(2013)]%
        {sorva2013review}
\bibfield{author}{\bibinfo{person}{Juha Sorva}, \bibinfo{person}{Ville Karavirta}, {and} \bibinfo{person}{Lauri Malmi}.} \bibinfo{year}{2013}\natexlab{}.
\newblock \showarticletitle{A Review of Generic Program Visualization Systems for Introductory Programming Education}.
\newblock \bibinfo{journal}{\emph{ACM Transactions on Computing Education (TOCE)}} \bibinfo{volume}{13}, \bibinfo{number}{4} (\bibinfo{year}{2013}), \bibinfo{pages}{1--64}.
\newblock
\href{https://doi.org/10.1145/2490822}{doi:\nolinkurl{10.1145/2490822}}


\bibitem[Walkington and Bernacki(2019)]%
        {walkington_personalizing_2019}
\bibfield{author}{\bibinfo{person}{Candace Walkington} {and} \bibinfo{person}{Matthew~L. Bernacki}.} \bibinfo{year}{2019}\natexlab{}.
\newblock \showarticletitle{Personalizing {Algebra} to {Students}’ {Individual} {Interests} in an {Intelligent} {Tutoring} {System}: {Moderators} of {Impact}}.
\newblock \bibinfo{journal}{\emph{International Journal of Artificial Intelligence in Education}} \bibinfo{volume}{29}, \bibinfo{number}{1} (\bibinfo{date}{March} \bibinfo{year}{2019}), \bibinfo{pages}{58--88}.
\newblock
\showISSN{1560-4306}
\href{https://doi.org/10.1007/s40593-018-0168-1}{doi:\nolinkurl{10.1007/s40593-018-0168-1}}


\end{thebibliography}

\end{document}